\documentclass[12pt]{article}

\usepackage{appendix}
 \usepackage{amssymb}
\usepackage{graphicx}
\usepackage{ifpdf}
\usepackage[fleqn]{amsmath}
\usepackage{bm}

\usepackage{authblk}

\newcommand{\lx}[1]{{\mathrm{#1}}}  		
\newcommand{\vc}[1]{\bm{#1}}  			
\newcommand{\order}{\mathcal{O}}			
\newcommand{\Aver}{\mathbb{E}}			
\newcommand{\vsb}[2]{{#1}_{#2}}  	 
\newcommand{\sts}{\Delta t_l}			
\newcommand{\Ph}{\hat{P}}			
\newcommand{\Pl}{\hat{P}_l}			
\newcommand{\Plnl}{\hat{P}_l^{N_l}}			
\newcommand{\dPl}{\delta\hat{P}_l}			
\newcommand{\dPlN}{\delta\hat{P}^{N_l}_l}			
\newcommand{\dPLN}{\delta\hat{P}^{N_L}_L}			
\newcommand{\PL}{\hat{P}_L}			
\newcommand{\PLnL}{\hat{P}_L^{N_L}}			

\newcommand{\MSE}{\lx{MSE}}			
\newcommand{\Var}{\lx{Var}}			
\newcommand{\vth}{v_{\lx{th},b}}			

\newcommand{\pdif}[2]{\frac{\partial {#1}}{\partial {#2}}} 
\newcommand{\eqn}[1]{(\ref{#1})} 

\newcommand{\mm}{\mathcal{M}}			

\newcommand{\cii}{i}
\newcommand{\cjj}{j}
\newcommand{\ckk}{k}
\newcommand{\cll}{m}
\newcommand{\rii}{k}

\date{}

\title{Multilevel Monte Carlo simulation of Coulomb collisions}
\author[1]{M.S. Rosin\thanks{Corresponding author: msr35@math.ucla.edu}}
\author[1]{L.F. Ricketson}
 \author[2]{A.M. Dimits}
\author[1,3]{R.E. Caflisch}
\author[2]{B.I. Cohen}
 \affil[1]{Mathematics Department, University of California at Los Angeles, Los Angeles, CA 90036}
  \affil[2]{Lawrence Livermore National Laboratory, L-637, P.O. Box 808, Livermore, CA 94511-0808}
  \affil[3]{Institute for Pure and Applied Mathematics, UCLA, Los Angeles, CA 90095}

\begin{document}
\maketitle

\begin{abstract}
We present a new, for plasma physics, highly efficient multilevel Monte Carlo numerical method for simulating Coulomb collisions. The method separates and optimally minimizes the finite-timestep and finite-sampling errors inherent in  the Langevin representation of the Landau-Fokker-Planck equation. It does so by  combining 
multiple solutions to the underlying equations with varying numbers of timesteps. For a desired level of accuracy  $\varepsilon$, the computational cost of the method is $\order(\varepsilon^{-2})$ or $\order(\varepsilon^{-2} (\ln \varepsilon)^2)$, depending on the underlying discretization,  Milstein or Euler-Maruyama respectively. This is to be contrasted
with a cost of $\order(\varepsilon^{-3})$ for direct simulation Monte Carlo or binary collision methods. We successfully demonstrate the method with a classic beam 
diffusion test case in 2D, making use of the L\'evy area approximation for the correlated Milstein cross terms, and generating a computational saving of a factor of $100$ for $\varepsilon = 10^{-5}$. We discuss the importance 
of the method for problems in which collisions constitute the computational rate limiting step, and its limitations. 

\end{abstract}

\section{Introduction}
In many regimes of practical importance, Coulomb collisions are an integral part of any accurate plasma description. For highly collisional systems, they are 
essential for closing the moment hierarchy of the kinetic equation and deriving microphysical expressions for the fluid transport coefficients. For marginally
collisional systems with order one Knudsen numbers, they play an important role in the dynamics, for example in tokamak edge plasmas \cite{koh2012bootstrap, park2010plasma}, inertial confinement fusion \cite{cohen2006effects}, and astrophysics \cite{bosch1992collision}. For weakly collisional, 
or `collisionless' systems, they regulate nonlinear phase space cascades of generalized energy and entropy \cite{Schek_09_cascade,tatsuno2009nonlinear}, and can be used to understand and control grid errors in numerical simulations.

This paper presents a new (for plasma physics applications) 
 accurate and efficient multi-(time-) level computational method for collisional kinetic problems, and is especially useful for systems in the low Knudsen number, i.e. highly collisional, regime. The method leverages a stochastic differential equation (SDE), or Langevin, approach to  solving the kinetic equation particle-wise.   It then combines the solutions using the multilevel Monte Carlo (MLMC) scheme,
initially developed for applications in financial mathematics \cite{giles2008multilevel} and now used in a wide variety of disparate areas \cite{giles2013review}.

 The MLMC method generates computational savings by separating and independently minimizing the finite-timestep and finite-sampling errors
inherent in any numerical SDE solver. Analogous to deterministic multigrid methods \cite{bungartz2004sparse}, the method builds a solution  
calculated from a weighted sum, over  different `levels' $l$, of successively refined building-block solutions obtained by direct methods like, for example, the Euler-Maruyama or Milstein discretizations. The so called `strong convergence' properties of these direct schemes determine the efficiency of the MLMC scheme in terms of a  \emph{global} error bound $\varepsilon$ in expectation, over all particles, of the time-integrated solution 
of the underlying SDE.

The solutions returned by the MLMC method are accurate approximations of the mean, with respect to the particle distribution function $f$, 
of any Lipschitz `payoff' function $P$ of the generalized phase space coordinates. This can include 
 the physically important macroscopic velocity moments of $f$, such as the density, fluid velocity, and temperature, that are governed by the moments of the underlying kinetic equation.  For example,   in the  case of a homogeneous, force-free, collisional plasma, the fluid velocity  is governed by the first 
moment equation
\begin{equation}
n \pdif{\vsb{u}{\cii}}{t} = \vsb{R}{\cii},
\nonumber
\end{equation}
where $n = \int f \lx{d}^3 \vc{v}$, $\vsb{u}{\cii} = \int f \vsb{v}{\cii} \lx{d}^3 \vc{v}$ and $\vsb{R}{\cii} = \int (\partial f/\partial t)_\lx{coll}\, \vsb{v}{\cii} \lx{d}^3 \vc{v}$ are the macroscopic density, fluid velocity and mean collisional 
transfer of momentum, respectively. Here $t$ is time, $\vc{v}$ is the particle velocity with components $\vsb{v}{\cii}$, and $(\partial f/\partial t)_\lx{coll}$ is the Landau-Fokker-Planck 
collision operator \cite{landau1936kinetic}. Unlike other approaches based on solving derived fluid equations, the MLMC method  does 
not rely on collisional closures or \emph{ad hoc} truncation schemes. The macroscopic solutions accurately reflect the underlying microscopic dynamics because the kinetic equation is solved directly.

The advantages of the MLMC method should be considered within the broader context of numerical collision methods for kinetic problems: particle-based, hybrid, and continuum methods  \cite{caflisch1998monte, thomas2012review}. Each has its own merits. Particle based methods are simple, direct and converge at a rate independent of the number of dimensions, 
but carry a stochastic error that depends on the number of simulation particles as $\order\left(N^{-1/2}\right)$. Hybrid 
methods are versatile and efficient, but only lead to efficiency gains for partially thermalized systems \cite{ricketson2013entropy}. Continuum methods are 
deterministic, but scale poorly with increased total (velocity plus spatial) dimension and lack the robustness of particle methods. They must also respect stability and CFL-like constraints 
on their discretization - even in the absence of mean fields.

For Monte Carlo simulations (pure particle and particle-based hybrid methods), binary collisions, for example the methods of Takizuke and Abe, and Nanbu, are a popular option \cite{takizuka1977binary, nanbu1997theory}. These  collision methods fall into a class of quasi-Maxwellian Boltzmann equations that have been shown to be no less accurate than $\order (\Delta t^{1/2})$ in terms of their global truncation error \cite{bobylev2013monte}. Related analytic and numerical studies confirm this lower bound, and these schemes have been argued to be as fast as $\order(\Delta t)$	 in the best case scenario \cite{bobylev2000theory, wang2008particle, dimarco2010direct}. The sampling error of the  methods,  governed by the Central Limit Theorem,  scales as $\order \left(N^{-1/2}\right)$. The Langevin-based or SDE description presents an alternative.

Existing computational Langevin collision models have largely focused on the lowest order `Euler-Maruyama' approximation to the
underlying Langevin equation.  Starting with Ivanov and Shvets \cite{ivanov1978use, ivanov1980methodEng} various collision 
models have been developed that evolve some subset of the particle's energy, pitch angle and azimuthal angle \cite{painter1993data, jones1996grid, manheimer1997langevin, dettrick1999monte, sherlock2008monte,lemons2009small, cohen2010time}. In their most basic forms, the `weak' convergence errors associated with  these schemes are, like optimal binary methods, $\order (\Delta t)$.  Some of the models also include advanced numerical techniques like grid-based schemes or schemes that use the Euler-Maruyama discretization as a building block in, for example, predictor-corrector schemes. Further extensions include self-consistent field models \cite{qiang2000self}, gyrokinetic applications \cite{velasco2012isdep}, and laser-plasma applications \cite{cadjan1997stochastic,cadjan1999langevin}. 

Beyond the Euler-Maruyama scheme, the next approximation in the hierarchy of higher order schemes is the `Milstein' scheme.  Its basic weak convergence 
error is also, like the Euler-Maruyama scheme,  $\order(\Delta t)$, but its strong convergence properties are improved. In one dimension, the Milstein terms are easy to implement \cite{lemons2009small, cohen2010time}. In higher dimensions, two or more, the complex statistics and statistical correlations in orthogonal Milstein terms prevent a simple description. Because collisions are a fundamentally multi-dimensional process in velocity space, even in reduced frameworks like gyrokinetics, this has been a major impediment for the application of higher-order Langevin methods in plasma physics.  However, recent work provides a simple, efficient approximation to the statistically correlated component of the orthogonal Milstein terms and a proof of concept demonstration of their use for Coulomb collisions \cite{dimits2013higher}. 

Existing SDE collision models and, in the best case, binary collision models have the same order of accuracy $\order(\Delta t)$. Both methods also have the same computational cost $\sim \order(\varepsilon^{-3})$, which comes from the product of a factor $\varepsilon^{-1}$ from the timestepping cost and a factor $\varepsilon^{-2}$ from the sampling cost (a result we derive in section \ref{sec:MLMC_main}).  This is to be contrasted with the cost of the MLMC method, which uses discretized SDEs as building blocks. 

The Euler-Maruyama MLMC scheme is $\order\left((\ln \varepsilon)^2/\varepsilon\right)$ faster than both SDE and binary methods, for the same level of accuracy. The Milstein MLMC 
method is even faster, offering a relative saving of $\order(1/\varepsilon)$, and is optimal amongst all discretizations  \cite{giles2008improved}. This paper provides a proof and demonstration of these results. 

The layout of this paper is as follows. In section \ref{sec:Langevin} we introduce the Langevin representation of the Landau-Fokker-Planck collision operator, and its basic numerical representation. In section \ref{sec:MLMC} we review the MLMC method of Giles that uses, as its building block, the basic numerical representation of the collision operator. In section \ref{Test} we 
present the results of the MLMC method as applied to a collisional relaxation problem. In section \ref{Discussion} we describe some limitations of the method and sketch 
some potential avenues for extending it.   Finally, in section \ref{Conclusion} we summarize and conclude.

\section{Coulomb-Langevin equations}\label{sec:Langevin}

\subsection{Formulation} \label{Formulation}

The starting point for most plasma collision models is the Landau-Fokker-Planck operator  \cite{landau1936kinetic}. This describes the effect of many small-angle collisions on the evolution of the phase-space test-particle distributions function $f_a \equiv f_a(t,  \vc{v})$ of the charged plasma species $a$
\begin{equation}
\pdif{f_a}{t}  = \pdif{f_a}{t}\big|_\lx{coll}\equiv - \pdif{}{\vsb{v}{\cii}}
\left(\left( \pdif{h}{\vsb{v}{\cii}}\right) f_a\right)
+ \frac{1}{2} \frac{\partial^2}{\partial \vsb{v}{\cii}\partial\vsb{v}{\cjj}} \left(\frac{\partial^2 g}{\partial\vsb{v}{\cii}\partial\vsb{v}{\cjj}}f_a\right),
\label{LFP}
\end{equation}
where  $t$ is time, $\vc{v}$ is velocity with components $\vsb{v}{\cii}$ and repeated indices are summed over. The Rosenbluth potentials $h, g$ \cite{rosenbluth1957fokker} are given by
\begin{align} 
(\partial^2/\partial \vsb{v}{\ckk}\partial \vsb{v}{\ckk})h &= -4 \pi \sum_b \Gamma (1+ m_a/m_b) f_b, \label{Rosenbluth_h}\\
(\partial^4/\partial \vsb{v}{\ckk}\partial \vsb{v}{\cll}\partial \vsb{v}{\ckk}\partial \vsb{v}{\cll}) g &= -8 \pi \sum_b \Gamma f_b,\label{Rosenbluth_g}
\
\end{align}
where $\Gamma = 4\pi q_a^2 q_b^2 \Lambda/m_a^2$, the sum is over the index $b$ of the plasma field-particle species $f_b$, mass is $m$,  charge is $q$, and $\Lambda$ is the Coulomb logarithm.

An alternative representation of the integro-differential Coulomb collision operator \eqn{LFP}-\eqn{Rosenbluth_g} is a drag-diffusion SDE for the 
random variable $\vc{v}$, describing the
same stochastic memoryless (Markov) process.   Under the assumption of white-noise forcing, the SDE description can be shown to be equivalent to the Fokker-Planck or forward Kolmogorov representation (Chapter 9.3, \cite{van1992stochastic}).  

Recasting the distribution function $f_a$ as a sum of delta-function particles, indexed by $\rii$ (henceforth repressed)
\begin{equation}
f_a(t, \vc{v})  = \sum_\rii \delta(\vc{v} - \vc{v}^\rii(t)),
\label{particle_dist}
\end{equation}
the particle velocities are governed by Newton's Second Law, which in the special case of \eqn{LFP}, corresponds to the SDE  \cite{ivanov1980methodEng, van1992stochastic}
\begin{equation}
d \vsb{v}{\cii} = \vsb{F}{\cii} dt + \vsb{D}{\cii \cjj} d\vsb{W}{\cjj}.
\label{Langevin}
\end{equation}
Here the total force is the sum of a deterministic drag force with coefficient $\vsb{F}{\cii}$ and a stochastic diffusion force with coefficient $\vsb{D}{\cii\cjj}$  and Wiener, or Brownian, process  $\vsb{W}{\cii}(t)$ with a normal probability density and variance 
\begin{equation}
\Aver\left[\left[ 
\vsb{W}{\cii}(t_2) - \vsb{W}{\cii}(t_1) 
\right]^2
\right]= |t_2 - t_1|,
\nonumber
\end{equation}
where $\Aver$ is the expectation. The Brownian motions are independent for each particle and component of the velocity. See Table \ref{symbol_table} for a summary of notation.

In Cartesian coordinates, and adopting an `Ito interpretation' \cite{kloeden1992numerical}, $\vsb{F}{\cii}$ and $\vsb{D}{\cii\cjj}$ are
related to \eqn{LFP} by 
\begin{align}
\vsb{F}{\cii} &= (\partial/\partial \vsb{v}{\cii}) h, \nonumber \\
\vsb{D}{\cii\cjj} &=\left[(\partial^2/\partial \vsb{v}{\cii}\partial \vsb{v}{\cjj})g\right]^{1/2}, \nonumber
\end{align}
and when $f_a$ is in equilibrium, i.e. a Maxwellian with the same temperature and flow velocity as $f_b$, $\vsb{F}{\cii}$ and $\vsb{D}{\cii\cjj}$ are
also related to each other by 
the Einstein\footnote{In the general sense, `Einstein relations' express the balance of deterministic and diffusive fluxes in a Fokker-Planck type equation satisfied in equilibrium. The work of Einstein described the positional motion of particles suspended in a fluid (undergoing Brownian motion), so
that the discussion there was of fluxes crossing notional boundaries in configuration space. In this paper, the relevant fluxes are across boundaries 
in velocity space.} relations \cite{einstein1905Brownian}:
\begin{align}
\vsb{F}{\cii} f_a + \frac{\partial}{\partial \vsb{v}{\cjj}} \left( \vsb{D}{\cii\cjj} f_a \right) = 0. \label{Einstein}
\end{align}

In curvilinear coordinate systems or for other stochastic calculuses, e.g. the `Stratonovich interpretation', $\vsb{F}{\cii}$ and $\vsb{D}{\cii\cjj}$ can appear as mixed coefficients of the drag and diffusion terms in \eqn{Langevin} \cite{kloeden1992numerical}.

Without loss of generality, macroscopic forcing - electromagnetism, gravity, model terms - can be included in this formulation. Either directly in the coefficient 
of the deterministic drag term, or via an operator splitting procedure. Once it has been numerically discretized, \eqn{Langevin} 
presents a simple method for including collisions or other stochastic processes in particle-in-cell codes.  The method can also be applied to classes of stochastic kinetic systems more general than plasmas. 

\subsection{Numerical discretization}\label{sec:discrete}

\begin{figure}
    \centering
  \includegraphics[width=0.47 \textwidth]{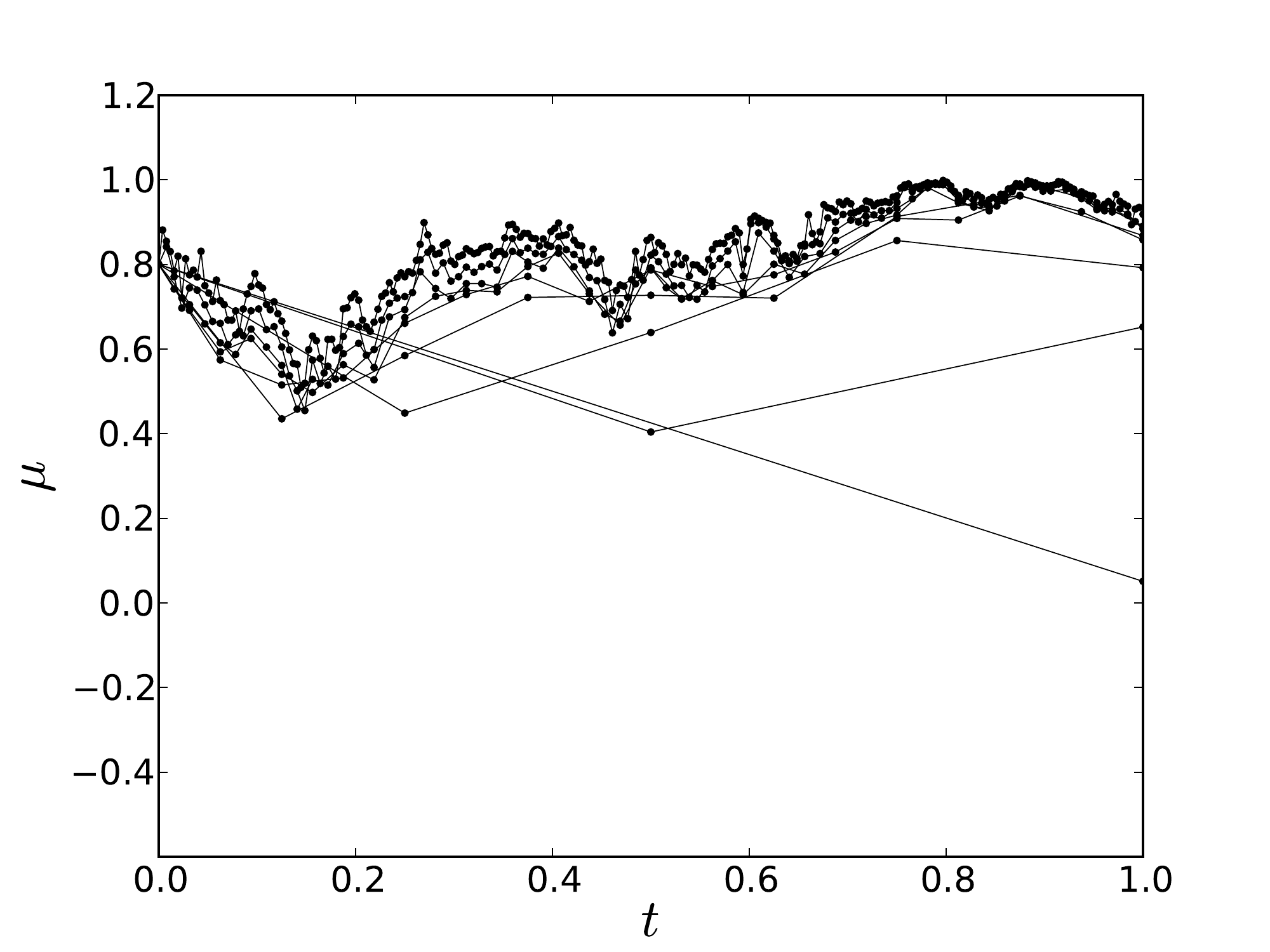}
  \includegraphics[width=0.47 \textwidth]{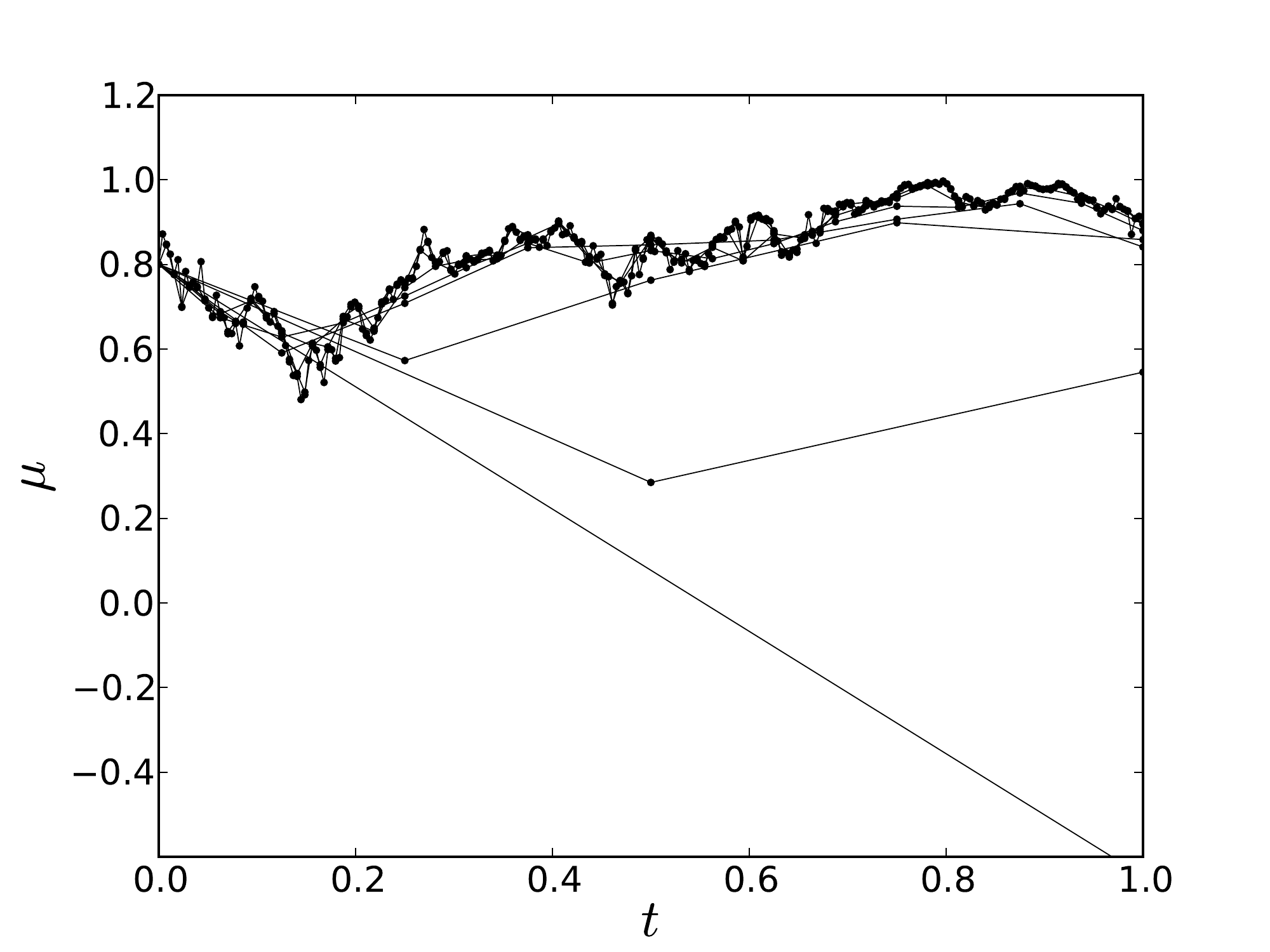}
  \caption{ Collisional evolution of velocity-space coordinate $\mu$ for a single particle using the Euler-Maruyama integration scheme (left) and Milstein scheme (right). 
  Results are generated from \eqn{Deltav_SPC} and  \eqn{Deltamu_SPC} using successively compounded timesteps $\sts = T 2^{-l}$ for $l = 0,1,2\ldots 8$, and the same underlying Brownian path. The rapid convergence of the Milstein results, with increasing $l$, are indicative of the scheme's improved strong convergence properties relative to the  Euler-Maruyma scheme.  Pairs of paths with $l, l-1$ are combined in the MLMC scheme to  estimate $\Aver[\mu]$. 
  \label{Fig:Path}}
\end{figure}

In general, solutions to \eqn{Langevin} at time $T$, $\vc{v}(T)$ must be obtained numerically. Discretization can be achieved by an iterative (stochastic)-Taylor expansion in the finite timestep $\sts = 2^{-l} T$.

The simplest integration  scheme is the Euler-Maruyama scheme, Fig. \ref{Fig:Path}. It is
\begin{equation}
\Delta \vsb{v}{\cii} = \vsb{F}{\cii} \sts  + \vsb{D}{\cii\cjj} \Delta \vsb{W}{\cjj}, 
\label{Euler}
\end{equation} 
where $\Delta \vsb{v}{\cii} =\vsb{v}{\cii}(t + \sts) - \vsb{v}{\cii}(t)$,  $\Delta \vsb{W}{\cjj} = \vsb{W}{\cjj}(t + \sts) - \vsb{W}{\cjj}(t)$, and under the Ito interpretation, 
the coefficients $\vsb{F}{\cii}, \vsb{D}{\cii\cjj}$, and their derivatives are to be evaluated at time $t$. Solutions to \eqn{Langevin} obtained using schemes like \eqn{Euler} and its higher order extensions, are said to be obtained \emph{directly}, or using \emph{single level}  estimates, i.e. $l = \rm{constant}$.

The weak and strong convergence properties of a direct scheme, like \eqn{Euler}, can be defined in terms of its weak and strong errors  \cite{kloeden1992numerical, dimits2013higher}. When solving for $\vc{v}(T)$ these are, respectively, given by:
\begin{align}
\varepsilon_W(\vc{v},T,\Delta t) &= \left|
\Aver \left[\vc{v}(T) \right] -
\Aver\left[\vc{v}_{l}(T)\right] 
\right|,\nonumber\\
\varepsilon_S (\vc{v},T,\Delta t) &= \Aver \left[| \vc{v}(T) - \vc{v}_l(T)|^2\right]^{1/2},\nonumber
\end{align}
where $\vc{v}_l$ is the solution to \eqn{Langevin} obtained using the finite timestep $\sts$. 
When solving for some Lipschitz function of $\vc{v}(T), P[\vc{v}(T)]$, the definition of the weak error (although not the strong error) must be generalized, so that \cite{kloeden1992numerical}:
\begin{align}
\varepsilon_W(P(\vc{v}),T,\Delta t) &= \left|
\Aver \left[P(\vc{v}(T)) \right] -
\Aver\left[P(\vc{v}_{l}(T))\right] 
\right|,\label{weak}\\
\varepsilon_S (P(\vc{v}),T,\Delta t) & = \Aver \left[| \vc{v}(T) - \vc{v}_l(T)|^2\right]^{1/2}. \label{strong} 
\end{align}
For single-valued initial conditions for the random variables, the expectations are over Brownian paths $W_i$ only. For multi-valued initial conditions, expectations are over both initial conditions and Brownian paths. 

A scheme is said to converge weakly with $\order\left(\sts^\alpha\right)$ if $\varepsilon_W \leq c \sts^\alpha$, and strongly with $\order(\sts^{\beta})$ if $\varepsilon_S \leq c \sts^{\beta}$ as $\sts\to0$, where the $c$'s are (different) constants. While strong convergence is a straightforward generalization of deterministic numerical convergence, it is rarely of practical importance. In general, 
it is the weak convergence properties of a numerical SDE scheme that dictates its utility. 

\begin{figure}
  \centering
  \includegraphics[width=0.49\textwidth]{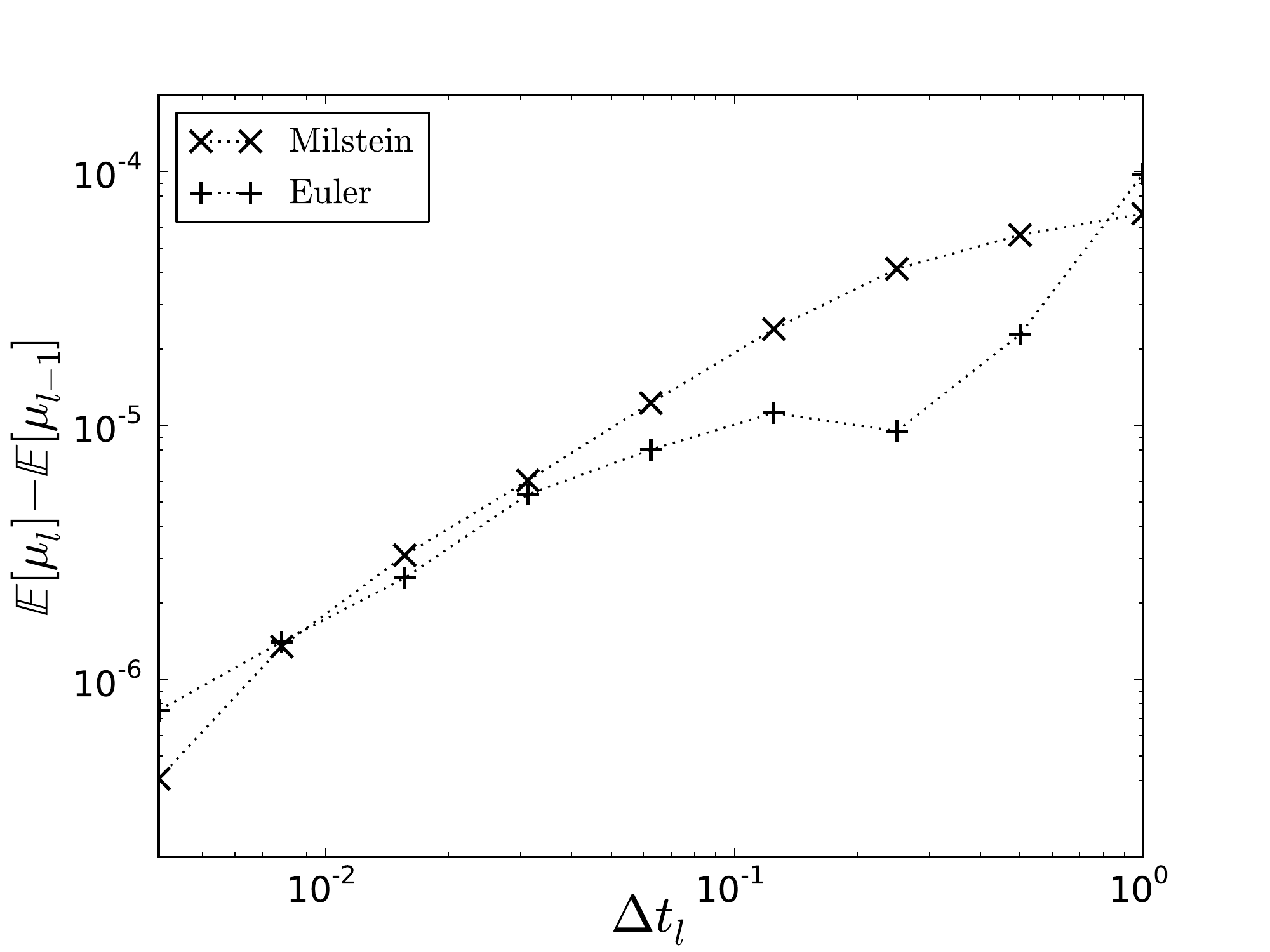}
    \includegraphics[width=0.49\textwidth]{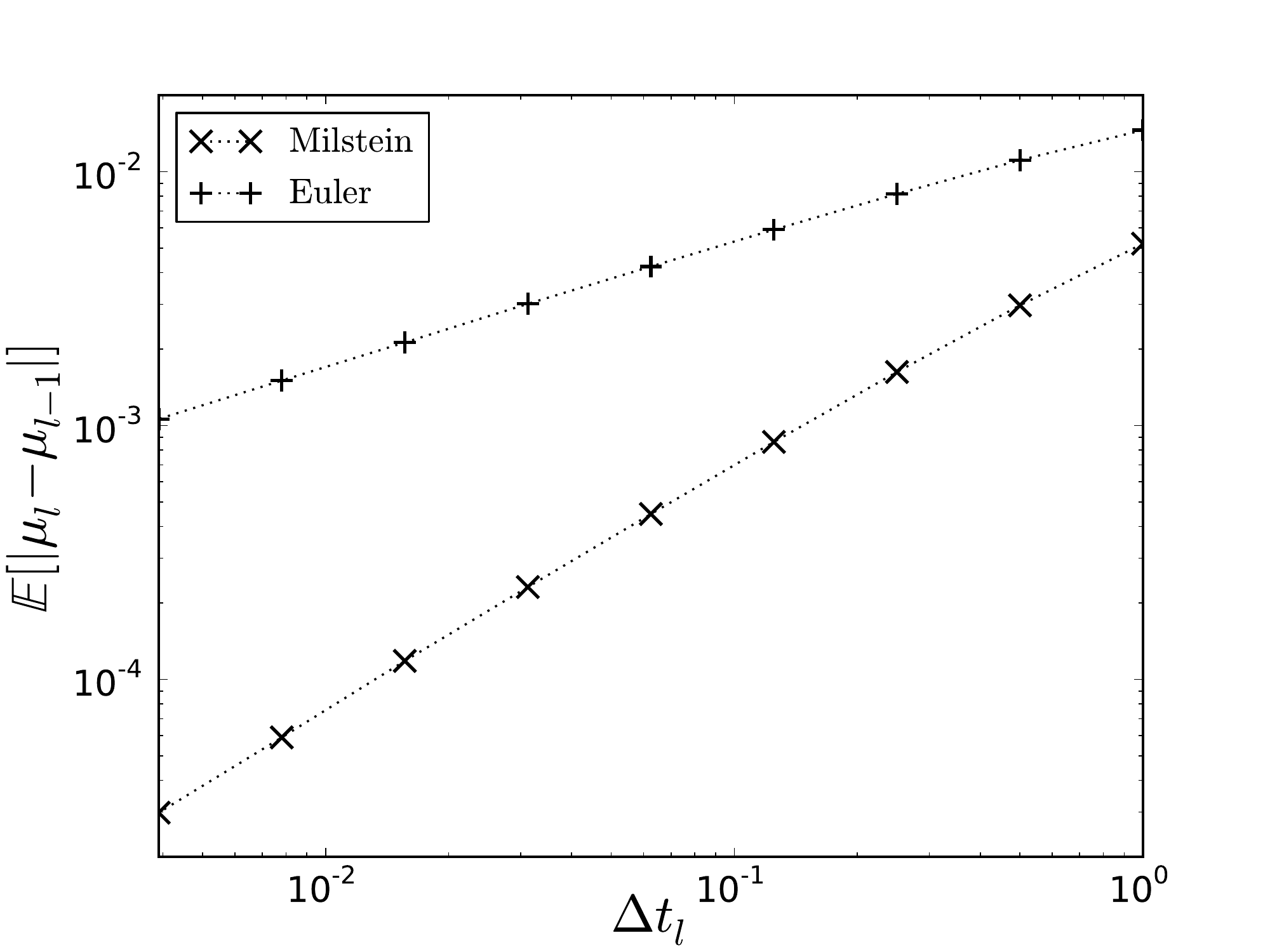}
  \caption{ Weak (left) and strong (right) scaling properties of the Euler-Maruyama and Milstein schemes for the $\mu$ component of the velocity calculated from $10^8$ samples. While both schemes have the same order of weak convergence, $\order(\sts)$, their strong convergence properties differ, \eqn{EulerScaling} and \eqn{MilsteinScaling}. The Milstein scheme convergence strongly as $\order (\sts)$, relative to the Euler-Maruyama scheme $\order ( \sqrt{\sts})$.  
  \label{Fig:Strong}}
\end{figure}

For the Euler-Maruyama scheme \eqn{Euler}, the convergence properties are
 \begin{equation} \label{EulerScaling}
P(\vc{v})- P(\vc{v}_{l})   = \left\{
     \begin{array}{llc}
 & \varepsilon_W \sim       \order(T \sts)  &- \quad \textrm{Weak Euler scaling,} \\
&\varepsilon_S \sim       \order(\sqrt{T \sts} ) &-  \quad \textrm{Strong Euler scaling,} 
     \end{array}
   \right. 
    \end{equation}
    so $\alpha = 1$ and $\beta = 1/2$, as shown in Fig. \ref{Fig:Strong}.

The next scheme in the hierarchy of Taylor expansions of \eqn{Langevin} is the first order Milstein approximation, also shown in Fig. \ref{Fig:Path}. It is  \cite{kloeden1992numerical, giles2012antithetic} 
\begin{equation}
\Delta \vsb{v}{\cii} =\vsb{F}{\cii} \sts  + \vsb{D}{\cii\cjj} \Delta \vsb{W}{\cjj} + \frac{1}{2} \vsb{D}{\cll \cjj}\pdif{\vsb{D}{\cii\cjj}}{\vsb{v}{\cll}} \left(\Delta\vsb{W}{\cjj}^2 - \sts\right) + 
\sum_{\cjj\neq \ckk} \vsb{D}{\cll\ckk}\pdif{\vsb{D}{\cii\cjj}}{\vsb{v}{\cll}} \vsb{A}{\ckk\cjj},
\label{Milstein}
\end{equation} 
where $\sts$ arising in the third term comes from the quadratic variation of a stochastic random variable, and $\vsb{A}{\ckk\cjj}$ is the off-diagonal `area integral' cross term given by 
\begin{equation}
\vsb{A}{\ckk\cjj} = \int^{t+\sts}_t  \!\!\! \left[ \vsb{W}{\cjj}(s) - \vsb{W}{\cjj}(t) \right]d \vsb{W}{\ckk}(s) = \int^{t+\sts}_t  \!\!\! d\vsb{W}{\ckk}(s) \int^{s}_t d\vsb{W}{\cjj}(\eta). 
\label{Levy}
\end{equation}

The area integrals $\vsb{A}{\ckk\cjj}$ are  non-Guassian random numbers that are closely related to the so-called `L\'evy areas' $\vsb{L}{\ckk\cjj} =(\vsb{A}{\cjj\ckk}-  \vsb{A}{\ckk\cjj})/2$, and are correlated with the Brownian motions $\vsb{W}{\ckk}$ and $\vsb{W}{\cjj}$ \cite{levy1951wiener}.  Numerically, sampling $\vsb{A}{\ckk\cjj}$ in a computationally efficient manner is technically challenging  \cite{kloeden1992numerical, gaines1997variable, gaines1994random}.  However, recently Dimits et al.  \cite{dimits2013higher} have developed a simple new approximate method for sampling $\vsb{A}{\ckk\cjj}$ in two dimensions. The method is simple to implement, inexpensive, accurate, and relies on the joint probability density function of the area integrals only. Using the prescriptions outlined in \cite{gaines1994random} and \cite{mezzadri2006generate}, it is expected that this two dimensional area integral can be used to generate the $D(D-1)/2$ non-independent area integrals  that arise in higher velocity-space dimensions D. 

The weak and strong convergence properties of the first-order Milstein scheme \eqn{Milstein} are 
 \begin{equation} \label{MilsteinScaling}
P(\vc{v})- P(\vc{v}_{l})     = \left\{
     \begin{array}{llc}
    & \varepsilon_W \sim       \order(T \sts) & - \quad \textrm{Weak Milstein scaling,}\\ 
    & \varepsilon_S \sim        \order(\sqrt{T} \sts ) & -  \quad \textrm{Strong Milstein scaling,} \\
     \end{array}
   \right. 
    \end{equation}
so $\alpha = 1$ and $\beta =1$, and, therefore, \eqn{Milstein} is superior to \eqn{Euler} only in its strong convergence properties, Fig. \ref{Fig:Strong}.

In the context of plasma physics, it is weak convergence that is typically important in simulating collisions \emph{directly} using  schemes like \eqn{Euler} and \eqn
{Milstein}. This is because plasmas are many particle systems in which it is the summed distribution $f$, as opposed to the individual particles, that are important. In other 
words, particle identity, which is incorporated into the strong error, is unimportant in constructing and evolving the distribution function.  

However, as we show in Section \ref{sec:MLMC}, it is the strong convergence properties of the underlying scheme that determines the computational efficiency of Giles' MLMC scheme. This is an instance of strong convergence being relevant to plasma physics. The MLMC method is significantly more efficient than direct methods, and especially so when used in conjunction with an underlying scheme with higher-order strong convergence. Quantitatively, the relationship between error, efficiency and computational cost, can be understood as follows.

\subsection{Efficiency and computational cost}\label{Sec:DirectCost}

\begin{table}
\centering
\begin{tabular}{ |l|l| }
  \hline
  \multicolumn{2}{|c|}{Commonly used notation} \\
  \hline
  $\vc{v}, \vsb{v}{\cii}$ & Velocity vector, components  \\
  $\vc{v}^\rii, \vsb{v}{\cii}^\rii$ & $\rii$-th realization of $\vc{v}, \vsb{v}{\cii}$\\
  $\vc{v}_l$ &  $\vc{v}$ calculated with $\sts$  \\
  $v, \mu, \phi$ & Spherical components of $\vc{v}$\\
  $\hat{\vc{v}}, \hat{\vc{v}_l}$ & $\Aver[\vc{v}], \Aver[\vc{v}_l]$ \\
  $ P, P_l$ & $P(\vc{v}), P(\vc{v}_l)$\\
  $\hat{P},\hat{P}_l$ & $\Aver[P(\vc{v})], \Aver[P(\vc{v}_l)]$\\
 $ V_l$ &$\Var[P_l- P_{l-1}]$\\
  \hline
\end{tabular}
\caption{Roman sub- and superscripts, with the exception of $l,N_l,L, N_L$, are vector components and random variable realizations respectively.}
\label{symbol_table}
\end{table}

The expectation of the solution to \eqn{Langevin}, $\Aver[\vc{v}(T)]$, has two sources of error in its numerical realization.  A finite-timestep error 
that depends on $\sts$, and a finite-sampling error that depends on the number of samples $N$. The same is true of any 
function of the solution, for example, the average kinetic energy $\mathcal{K}$ of a collection of particles:
\begin{equation}
\mathcal{K} = \frac{1}{2} \frac{m}{n} \int f(T, \vc{v} ) |\vc{v}|^2 d^3\vc{v}\equiv \frac{1}{2} m\,\Aver[|\vc{v}(T)|^2],\label{kinetic}
\end{equation}
where the left and right hand interpretations of $\vc{v}$ are as in \eqn{particle_dist}, so $f$ obeys \eqn{LFP} and $\vc{v}(T)$ obeys \eqn{Langevin}.

Minimizing the error in the moments of $f$, for example \eqn{kinetic}, is a compromise between efficiency and accuracy. 
Let $P= P(\vc{v}(T))$ be some Lipschitz scalar function of $\vc{v}(T)$, let $P_l = P(\vc{v}_l(T))$ be its finite timestep approximation, and let $P_l^\rii = P(\vc{v}^\rii_l(T))$ be the $\rii$-th sample of the finite timestep approximation. For numerical schemes that employ discretizations like \eqn{Euler} or \eqn{Milstein} directly, we define
\begin{align}
\Ph &= \Aver[P] & \lx{with} \quad & N\to \infty, \sts \to 0, \label{true}  \\
\Pl &= \Aver[P_l] & \lx{with} \quad & N \to \infty, \sts = 2^{-l} T, \label{Plapprox}  \\
  \Plnl &= N_l^{-1}\sum\nolimits_{\rii=1}^{N_l} P^\rii_l& \lx{with} \quad & N = N_l, \sts = 2^{-l} T. \label{Plnlapprox} 
\end{align}
to be  the `true', finite-timestep, and finite-timestep finite-sampling approximations respectively, Table \ref{symbol_table}.  

Equations \eqn{true}-\eqn{Plnlapprox} are calculated from \eqn{Langevin} in two stages. First, applying some convergent integration scheme with $\sts \to 0$ for $\vc{v}$, or $\sts = 2^{-l} T$  a constant for $\vc{v}_l$. Second, applying $P$ and calculating the expectation by generating multiple samples, and then averaging over them with $N\to \infty$ for $\Ph$ or $\Pl$, and finite $N = N_l$ for $\Plnl$. 

An accurate  estimate $\Plnl$ of $\Ph$ is then one for which the mean squared error (MSE)
 \begin{align}
 \MSE \equiv & \Aver\left[\left(\Ph - \Plnl\right)^2\right] \nonumber \\
 =& \left( \Ph - \Pl \right)^2 + \Aver
 \left[ 
 \left(\Pl - \Plnl
 \right)^2
 \right],
 \label{MSEDfn}
 \end{align}
is small. The final equality follows from the fact that $\Aver[\Pl - \Plnl] = \Pl - \Plnl$ and $\Plnl$ is an unbiased estimate of $\Pl$

The size of the two terms in \eqn{MSEDfn} can be varied independently. The first depends on the weak convergence rate of the scheme $\order(\sts^\alpha)$, and is independent of $N$. The second depends on the number of samples $N=N_l$, and its size is independent of $\sts$.  Their associated sizes are
\begin{equation}
\left( \hat{P} - \hat{P}_l\right)^2 \lesssim c_1^2 \sts^{2 \alpha}, \qquad  
\Aver\left[  \left(\Pl - \Plnl
 \right)^2 \right] = \frac{\Var[P^\rii_l]}{N_l},
 \label{DirectError}
\end{equation}
where 
\begin{equation} \Var[P] \equiv \Aver[ (P - \Aver[P])^2]\nonumber
\end{equation} is the variance operator on a random variable, and $c_1$ is a constant. 

It follows that $\Plnl$ is accurate to within $\varepsilon$ of $\Ph$ if
\begin{equation}
\varepsilon^2 \geq \MSE \sim c_1^2 \sts^{2\alpha} + \frac{\Var[P^\rii_l]}{N_l}.
\label{scaling}
\end{equation}
The challenge in enforcing this bound is to do so as efficiently as possible.  

For \emph{direct} integration,  the computational cost $K$ of obtaining  $\Plnl$ is the product of the number of timesteps $T/\sts = 2^l$ and the number of samples $N = N_l$. It provides a simple measure of efficiency and is defined as 
\begin{equation}
K =   N_l \frac{T}{\sts} ,
\label{compcost}
\end{equation}
In practice, $K$ measures the number of times the collision integration routine must be called in a numerical simulation. 

To make the scheme as efficient as possible, we wish to minimize $K$ subject to  \eqn{scaling}. To do so, we will 
ensure both terms in \eqn{MSEDfn} are individually bounded, so their sum $\leq \varepsilon^2$.
Applying the method of Lagrange multipliers in case of equality in \eqn{scaling} yields expressions for the optimal $\sts$ and
$N_l$:
\begin{align}
\sts &\simeq \varepsilon^{1/\alpha} \; [c_1^2\, (2 \alpha + 1)]^{-1/2\alpha},\nonumber\\
N_l &\simeq \varepsilon^{-2} \; \left( 1 + \frac{1}{2 \alpha} \right) \Var[P^\rii_l],\nonumber
\end{align}
Direct substitution into \eqn{compcost} then reveals the optimal computational cost is
\begin{equation}
K_{\lx{opt}} \simeq \frac{ c_2}{\varepsilon^{(2+1/\alpha)}} \frac{1}{2 \alpha}\left( 1 + \frac{1}{2\alpha}\right)^{1 + 1/2\alpha} \Var[P^\rii_l],
\label{CostDirect}
\end{equation}
where $c_2$ is a constant.

It is important to note that for both the Euler-Maruyama and the first order Milstein schemes, \eqn{Euler} and \eqn{Milstein},  $\alpha = 1$. It follows that  $K_{\lx{opt}} \sim 
\order(\varepsilon^{-3})$. It is only by including higher order  terms that the weak error scaling, and therefore the optimal computational cost, can be improved. 

While the direct approach has the advantage of being conceptually simple, 
it is asymptotically inefficient. Minimizing the error using \emph{direct} methods requires both a large sample size and a small step size, which tends to over-resolve the problem. It is this inefficiency that is improved upon by the MLMC method.

\section{Multilevel Monte Carlo method}\label{sec:MLMC}

\subsection{Background}\label{sec:MLMC_main}

The computational cost of direct methods scales with their timestep resolution and expectation sample size.  The improved efficiency of the MLMC method, relative to the methods in Section \ref{sec:discrete}, comes from judiciously expending computational resources only when necessary. As initially described by Giles \cite{giles2008multilevel}, and reviewed in this section, the improved efficiency of the method is achieved by building an estimate of $\hat{P}$ from multiple solutions with varying timesteps $\sts = 2^{-l} T$, i.e. values of $l$, and expectations with varying   sample sizes $N_l$. For the coarsest level $l=0$, the Langevin equation is integrated with a single timestep, while for the finest level $l=L$, $2^L$ timesteps
are required.

The basic mechanism behind the method's improved efficiency can be understood 
as follows. For small values of $l$,  estimates are inexpensive to compute accurately, because only a few timesteps are required for each realization of the
numerical solution. In turn, for large values of $l$, where each integration is relatively expensive, only a few realizations are needed because the
finite-sampling error converges to zero as the strong error, assuming $\beta$ is positive.

\begin{figure}
\includegraphics[width=0.48\textwidth]{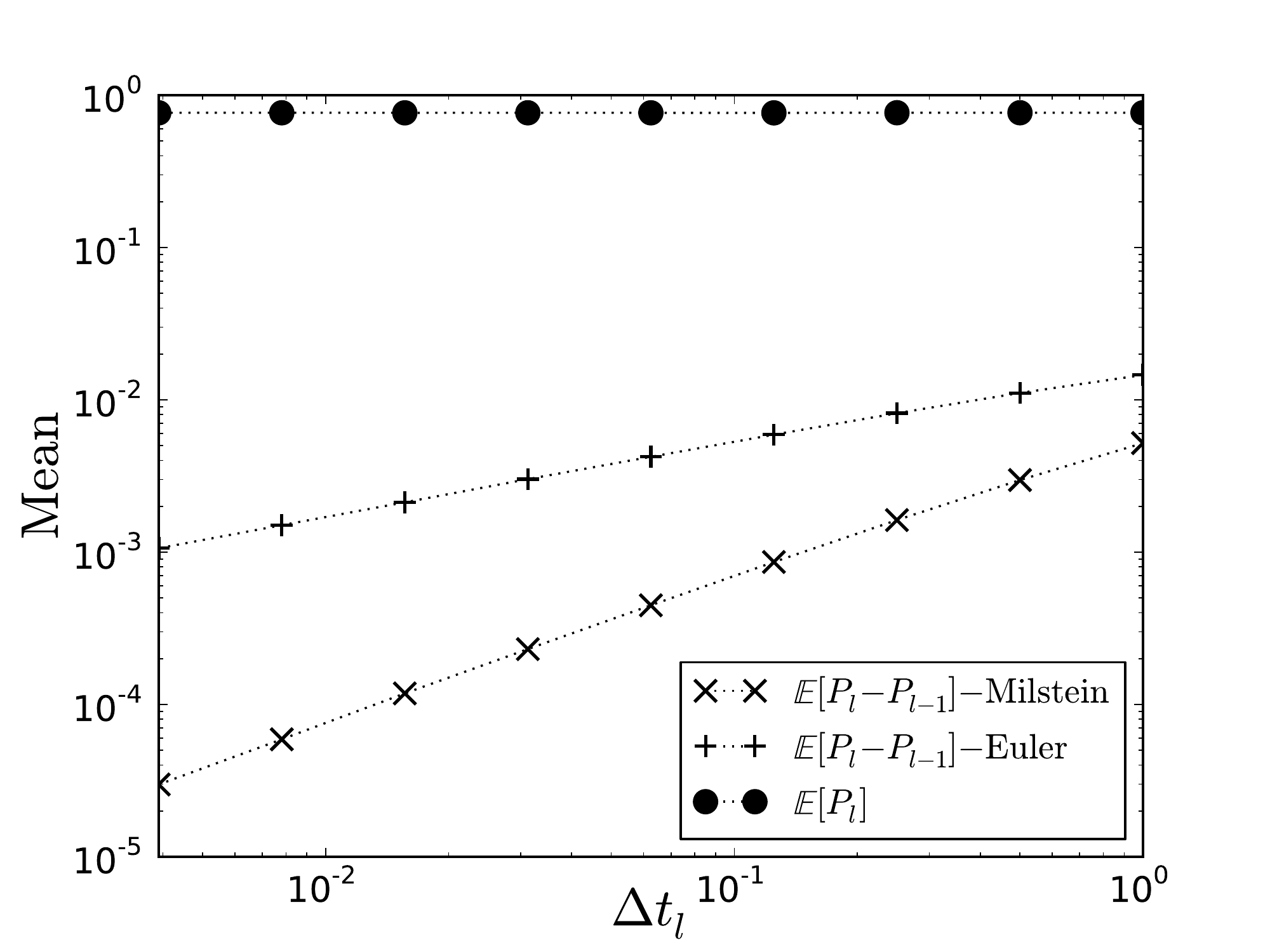}
\includegraphics[width=0.48\textwidth]{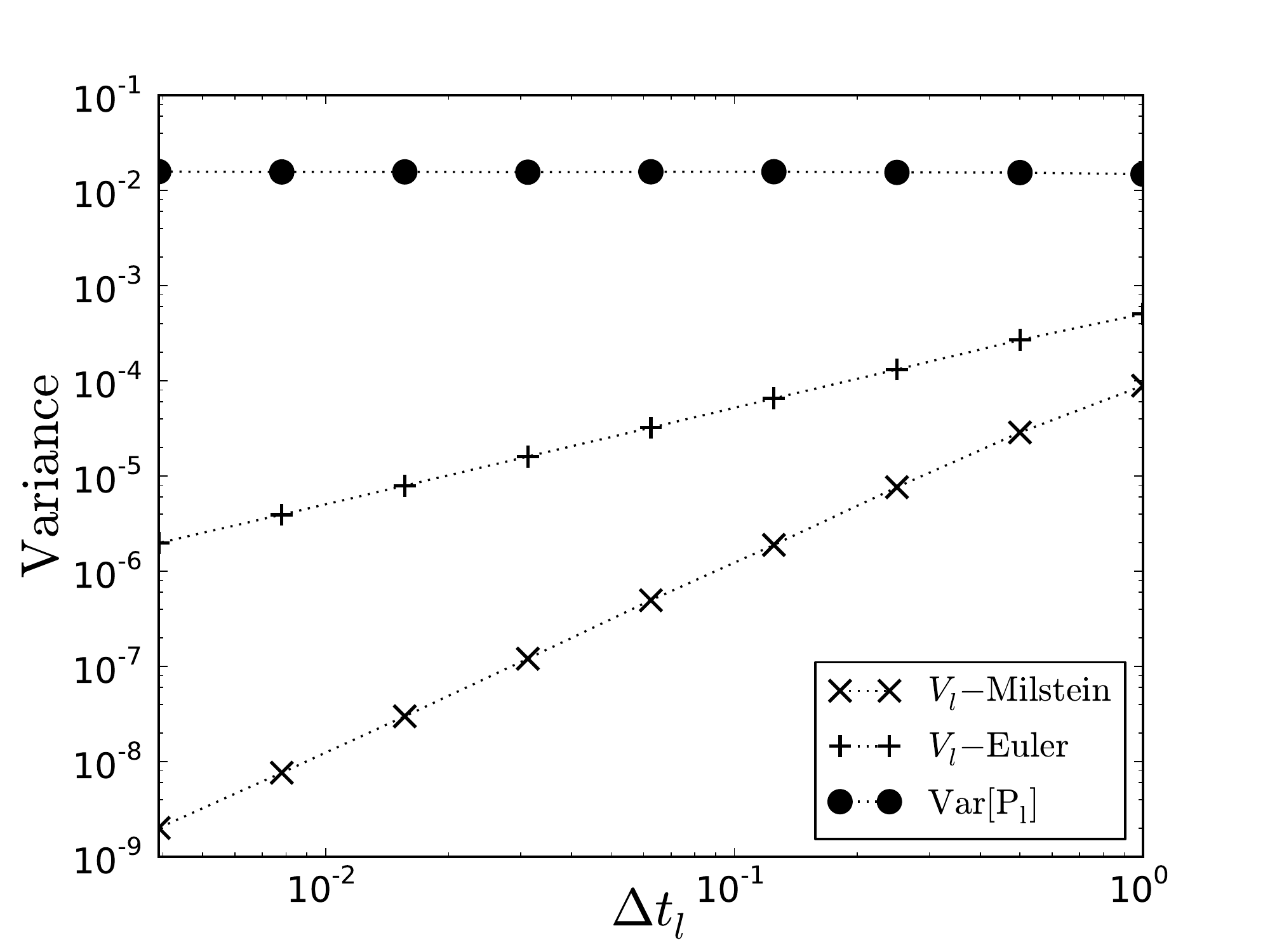}
  \caption{Mean (left) and variance (right) of the difference between levels for the Euler and Milstein schemes. The mean of the difference at level $l$ appears explicitly in \eqn{PayoffEst}. The variance in differences at $l$ is significantly less than that of a single level, which allows term in $\PLnL$ to be calculated efficiently using MLMC methods. Data is taken from the beam diffusion test case in Section \ref{Test} with payoff  $P = \mu$,  (delta-function) initial conditions  $v^* = 0.5, \mu^* = 0.8$, and final time $T = 0.02$. All quantities calculated in this figure are taken using $10^5$ samples. 
    \label{Fig:Mean}}
\end{figure}

From the linearity of the expectation operator, we have the following identity
\begin{align}
\hat{P}_L \equiv \Aver[P_L] &= \Aver[P_0] + \sum^L_{l=1}\Aver[P_l - P_{l-1}], \nonumber \\
&\equiv  \hat{P}_0 + \sum^L_{l=1} \dPl
\label{generalMLMC}
\end{align}
where $\hat{P}_0 = \Aver[P(\vc{v}_0)]$ is estimated using a single timestep and $\dPl \equiv \Aver[P_l - P_{l-1}]$.  Equation \eqn{generalMLMC} describes a
telescoping sum, where the contribution of each term decreases with increasing $l$, as shown in Fig. \ref{Fig:Mean}.

The finite sampling analogue of \eqn{generalMLMC} can  be obtained by generating $N_0, N_l$ samples of $\hat{P}_0,  \dPl$ and combining
them so 
\begin{equation}
\PLnL = \hat{P}_0^{N_0} + \sum_{l=1}^L \dPlN 
\label{PayoffEst}
\end{equation}
where 
\begin{align}
P_0^{N_0} &=  \frac{1}{N_0} \sum_{\rii=1}^{N_0} P^\rii_0,\\
\dPlN &= \frac{1}{N_l} \sum_{\rii=1}^{N_l}   (P^\rii_l - P^\rii_{l-1}),
\label{PayoffDiff}
\end{align}
are  unbiased estimates of $\hat{P}_0, \dPl$ respectively. 

In calculating each pair $P_l^\rii$ and $P_{l-1}^\rii$ that contributes to the sum in $\dPlN$, it 
is essential that the payoffs are constructed from the same underlying stochastic path and initial conditions. That is, for each contributing realization to $\dPlN$,  $P_{l-1}^\rii$ must be constructed by 
suitably coarsening $P_{l}^\rii$, or conversely, $P_{l}^\rii$ must be calculated by suitably refining  $P_{l-1}^\rii$. One coarsening method, including a prescription for the
multi-dimensional L\'evy areas,  is provided in section 4.3 of Dimits et al. \cite{dimits2013higher}. Paths and initial conditions for different realizations that contribute to $\dPlN$, and indeed different $\dPlN$'s and $\hat{P}_0^{N_0}$, can be calculated independently.

Equation \eqn{PayoffEst} returns a good estimate of $\hat{P}, \PLnL,$ if, for a reasonable computational cost, the total error is small.  Like the direct methods of Section \ref{Sec:DirectCost}, the finite-timestep contribution to the total error is governed by the weak convergence properties of the underlying scheme.  However, unlike direct methods and crucially for the MLMC method,
the finite-sampling, or variance, contribution
\begin{equation}
\Var[\PLnL] = \Var[\hat{P}^{N_0}_0] +  \sum_{l=1}^L \Var[\dPlN], 
\nonumber
\end{equation} is determined by the strong convergence properties of the underlying scheme.

As in \eqn{MSEDfn}, the mean square error is given by 
\begin{align}
\MSE &=  \left( \Ph - \PL \right)^2 + \Aver
 \left[ 
 \left(\PL - \PLnL
 \right)^2
 \right],
 \label{MSEDfn2}
 \end{align}
 which we wish to bound so that $\varepsilon^2 \geq \MSE$. Analogous to \eqn{DirectError}, the two terms are of size
\begin{equation}
\left( \hat{P} - \hat{P}_L\right)^2 \lesssim c_1^2 \sts^{2 \alpha}, \qquad  
\Aver\left[ 
 \left(\PL - \PLnL
 \right)^2
 \right] = \frac{V_0}{N_0} +  \sum_{l=1}^L \frac{V_l}{N_l},
 \label{MSE2}
\end{equation}
where $V_l \equiv \Var[P^\rii_l - P^\rii_{l-1}]$ and $V_0 = \Var[V_0^\rii]$ are the variances of a single sample. The variances of these 
samples  are related to those of the random variable $\dPlN$ by $\Var[\dPlN] \simeq V_l/N_l$ and $\Var[\hat{P}_0^{N_0}] \simeq V_0/N_0$.
For $l>0$, $V_l$ follows the strong convergence order of the underlying scheme \eqn{strong}:
\begin{equation}
V_l \lesssim c_3 \sts^{2\beta}.
\label{VlScaling}
\end{equation}
where $c_3$ is a constant.
We demonstrate this result numerically in Fig. \ref{Fig:Mean}, and note that \eqn{VlScaling}  dictates that the finite-sampling error in $\dPlN$ can be bounded using fewer and fewer samples $N_l$, as $l$ increases ($\sts$ decreases).  

From \eqn{MSEDfn2} and \eqn{MSE2}, it follows that $\PLnL$ is a good estimate of $\hat{P}$ if 
\begin{equation}
 \varepsilon^2 \geq \MSE \sim c_1^2 \sts^{2\alpha} + \frac{V_0}{N_0} +  \sum_{l=1}^L \frac{V_l}{N_l},
 \label{MLMCscaling}
\end{equation}
which has an associated computational cost of
\begin{equation}
K = \sum^L_{l=0} K_l \equiv  \sum^L_{l=0}  N_l\frac{T}{\sts}.
\label{CompCost2}
\end{equation}
The most efficient method for calculating $\PLnL$ is, again,  the one that minimizes $K$ subject to \eqn{MLMCscaling}. Unlike \emph{direct} methods, there are now
two new degrees of freedom over which to optimize: the total number of levels $L$, and the number of samples used for the
expectation at each level $N_l$. As in the previous section, the minimal $K$ will clearly occur when $\MSE = \varepsilon^2$ and so we approach the problem
by separately bounding the two terms in \eqn{MSEDfn2} as follows:
\begin{equation}
\left( \Ph - \PL \right)^2 = \frac{1}{2} \varepsilon^2, \qquad
\Aver
 \left[ 
 \left(\PL - \PLnL
 \right)^2
 \right] = \frac{1}{2} \varepsilon^2. 
 \label{condition}
\end{equation}

\begin{figure}
\includegraphics[width=0.48\textwidth]{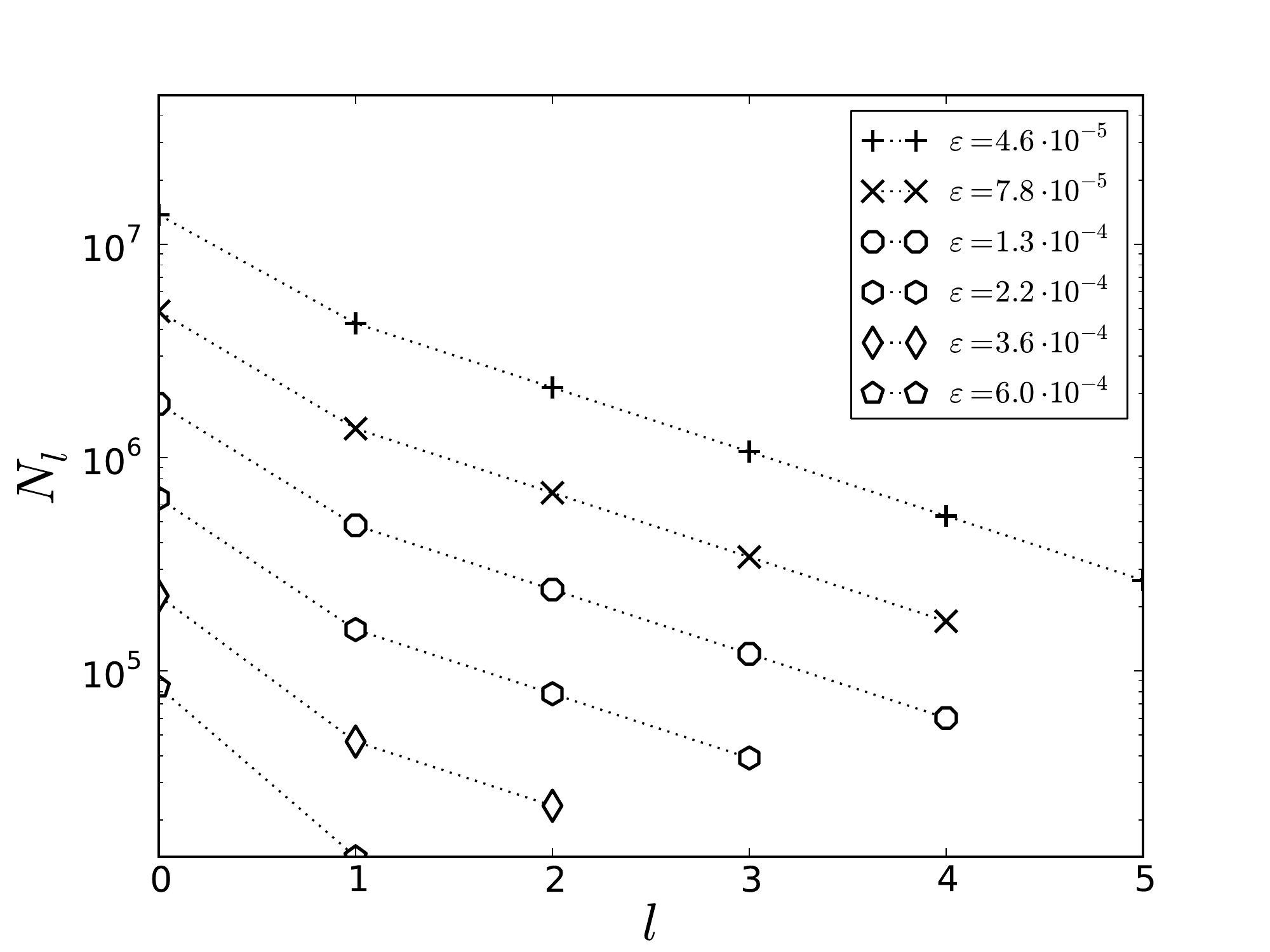}
\includegraphics[width=0.48\textwidth]{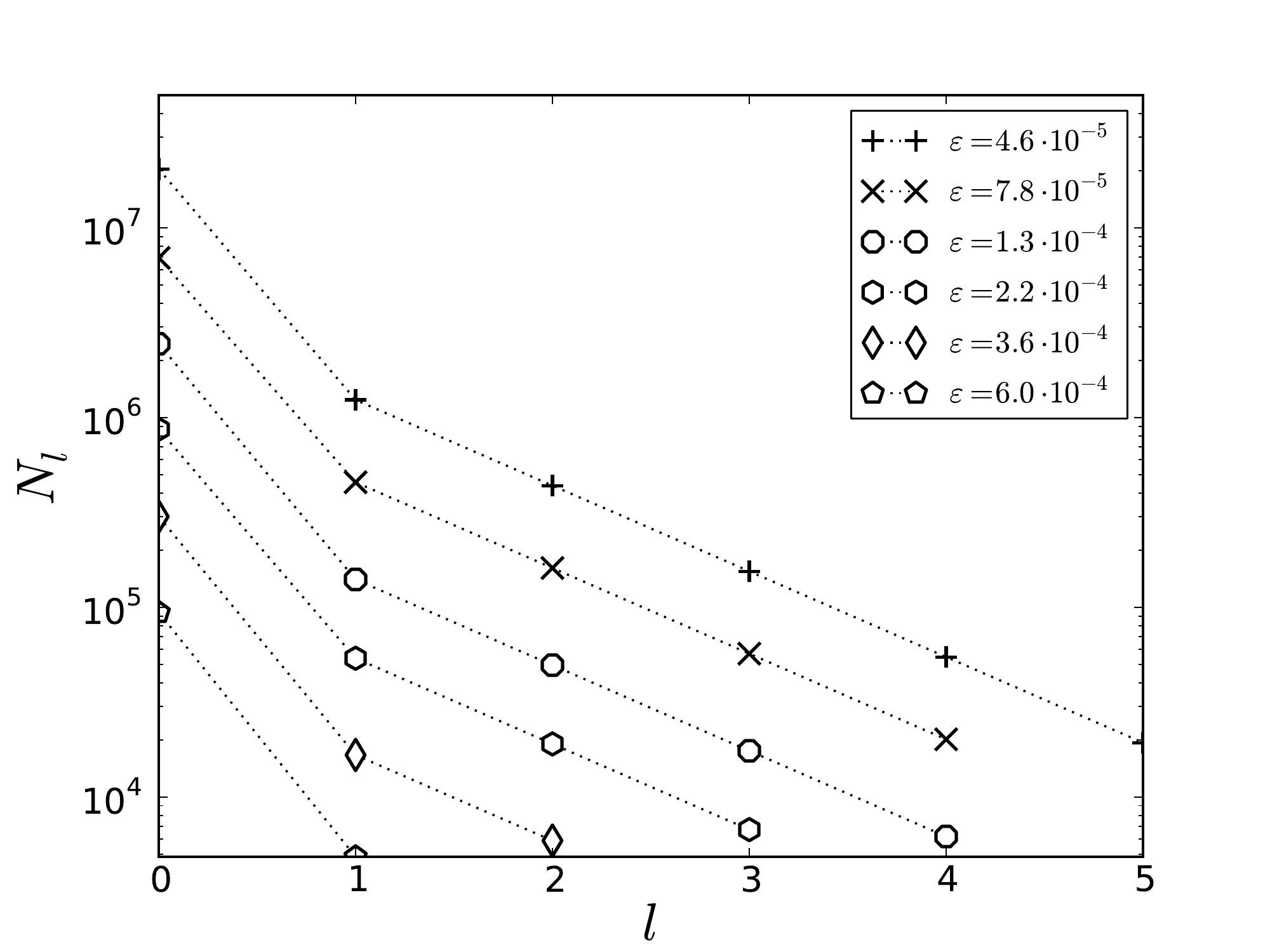}
  \caption{Samples $N_l$ at each level $l$ for the MLMC scheme with Euler (left) and Milstein (right) discretizations, \eqn{OptNumber}. The scaling 
  for levels $l\geq1$ is determined by the strong convergence properties of the underlying scheme, $N_l \sim 2^{-l \beta}$. The computational cost at each level $K_l = N_l 2^l$ is approximately constant for the Euler method, but decreases rapidly in the telescoping Milstein sum.  Parameters are the same as those used in Fig. \ref{Fig:Mean}.
    \label{Fig:Nl}}
\end{figure}

The first condition, along with \eqn{MLMCscaling} gives
\begin{equation}
L = \frac{1}{\alpha}
\frac{ \ln \left[ c_1 T^\alpha \sqrt{2} \varepsilon^{-1}\right]}
{\ln 2}.
\label{OptLevel}
\end{equation}
Considering this $L$ fixed, a Lagrange multiplier argument reveals the optimal efficiency is obtained when $N_l \sim \sqrt{V_l T 2^{-l}}$. Using this and 
the second condition in \eqn{condition}, the optimal number of samples at level $l$ is given by
\begin{equation}
N_l = \frac{\sqrt{V_l 2^{-(l+2)}}}{ \varepsilon^2}  \sum_{l=0}^L \sqrt{V_l 2^l}, 
\label{OptNumber}
\end{equation}
where \eqn{VlScaling} ensures that $N_l$ is a strictly decreasing function of $l$ as shown in Fig. \ref{Fig:Nl}. 

Now, combining  \eqn{VlScaling}, \eqn{CompCost2}, \eqn{OptLevel} and \eqn{OptNumber}, the optimal computational cost of the MLMC scheme is given by
\begin{equation}
K_{\lx{opt}} \simeq \frac{2 c_4  T^{(2 \beta -1)/2}}{\varepsilon^2} \left(\sum^L_{l=0} 2^{-l\,(2 \beta-1)/2}\right)^2,
\label{MLMCOpt}
\end{equation}
where $L = L(\varepsilon)$ is given by  \eqn{OptLevel}. In the case 
of $\beta = 1/2$, the sum in \eqn{MLMCOpt} scales  as $L\sim\ln \varepsilon$, whereas for $\beta >1/2$, the sum can be uniformly bounded. From this, the asymptotic cost of the MLMC method is $\order(\varepsilon^{-2} (\ln  \varepsilon)^2)$ for the Euler-Maruyama method and  $\order(\varepsilon^{-2})$ for the Milstein method.

These costs are to be contrasted with the total cost of direct and binary methods, for which $K$ is given by  \eqn{compcost}. As described in Section \ref{Sec:DirectCost}, the computational cost of these methods can be easily 
calculated by writing the requisite (so that the $\lx{MSE} \leq \varepsilon^2$) timestep $\sts$ and sample size $N$ in terms of $\varepsilon$ and substituting directly into \eqn{compcost}. For direct methods, the result of doing so is given by \eqn{CostDirect} so $K \sim \order\left(\varepsilon^{-(2+1/\alpha)}\right)$ for a general weak order-$\alpha$ scheme, and $K \sim \order(\varepsilon^{-3})$ for the widely used $\alpha =1$ direct Euler-Maruyama integration scheme. For binary methods, the analysis is identical\footnote{Note that binary collision algorithms pair particles into $N/2$ sets when performing collisions. This offers a relative saving of up to a factor of a half, compared to Langevin treatments  \cite{cohen2013grid}, although the constant factor does not affect the 
scaling properties of the algorithm.}.
The 
finite-timestep error is, at best,  $\order(\sts)$ and the finite-sampling error is $\order(N^{-1/2})$, so the requisite scalings of these two terms are $\sts \sim \order(\varepsilon)$ and $N \sim \order(\varepsilon^{-2})$ respectively. It follows that for the binary method, at best, $K \sim \order(\varepsilon^{-3})$ and, at worst, when the finite-timestep error is $\order(\sts^{1/2})$, the cost is $K \sim \order(\varepsilon^{-4})$.

The relative theoretically optimal costs of the 
various methods are therefore: 
 \begin{equation} \label{CostSummary}
K_\lx{opt}  = \left\{
     \begin{array}{ll}
       \order\left(\varepsilon^{-3}\right) & - \quad \textrm{Binary\;collisions}, \\
       \\
       \order\left(\varepsilon^{-(2 + 1/\alpha)}\right) & - \quad \textrm{General\;order-}\alpha\textrm{\;direct\;SDE}, \\
       \\
       \order\left(\varepsilon^{-2} (\ln \varepsilon)^2 \right) & -  \quad \textrm{MLMC\;with\;} \beta = 1/2, \\
	\\
        \order\left(\varepsilon^{-2}  \right) & -  \quad \textrm{MLMC\;with\;} \beta > 1/2. 
     \end{array}
   \right. 
    \end{equation}

In Fig. \ref{Fig:CostErr2} we consider the specific test case of the collisional relaxation of a monoenergetic, low-density beam, as described in Section \ref{Test}. The figure 
confirms that the cost scaling in \eqn{CostSummary} are accurate  for the $\alpha =1$ direct Euler method, and the Euler and Milstein multilevel schemes. It also 
shows that the  computational cost of the MLMC method is substantially less than that of the direct method.

\subsection{Numerical Implementation}\label{Implementation}

Equations \eqn{PayoffEst}-\eqn{PayoffDiff}, \eqn{OptLevel} and \eqn{OptNumber} provide a prescription for approximating $\hat{P}$ by $\PLnL$,
such that its $\MSE \leq \varepsilon^2$. There are two degrees of freedom in the MLMC scheme, $L$ and $N_l$, each influencing an associated finite-timestep and finite-sampling error. The constants that determine $L$ and $N_l$, such that \eqn{condition} is enforced, are $c_1$ and $V_L$ respectively. 

In the asymptotic limit of small timestep (large $l$), $c_1$ is the constant of proportionality between $\hat{P} - \hat{P}_l$ and $\sts$, as defined in
the \emph{weak} error \eqn{weak}. It can be calculated using Richardson extrapolation
\begin{equation}
|c_1| \simeq |c_1|_{(N)} \equiv \frac{|\hat{P}^N_{l} - \hat{P}^N_{l-1}|}{T^\alpha 2^{-l \alpha} | 1 - 2^{\alpha}|},
\label{cone}
\end{equation}
where $P^N_{l}, P^N_{l-1}$ are determined empirically by direct integration with the relevant discretization (Euler-Muruyama, first-order Milstein) and the integer $N\gg1 $ is large enough that the sampling error in $\hat{P}^N_l$ is small relative to the timestep error i.e. $\hat{P}^N_l \simeq \hat{P}_l$. This semi-equality can be checked, \emph{ex post facto}, by ensuring that 
\begin{equation}
1 \gg \frac{|c_1|_{(nN)} - |c_1|_{(N)}}{|c_1|_{(nN)}},
\nonumber
\end{equation}
for $n> 1$, also an integer. 

As for $c_1$, $V_L$, which influences the finite sampling error in the MLMC scheme, must be determined empirically. It can be estimated by taking
$N$ samples of $P_L - P_{L-1}$, and setting
\begin{equation}
V_L \simeq V_L^{N} \equiv \frac{1}{N} \sum_{\rii=1}^N \left( P^\rii_L - P^\rii_{L-1} \right)^2 - \frac{1}{N^2} \left(\sum_{\rii=1}^N \left(P^\rii_L - P^\rii_{L-1}\right) \right)^2, 
\label{VL}
\end{equation}
where each value of $P$ in the pair is generated from the same Brownian path and initial condition using the relevant discretization, and $N$ should be large enough to ensure good statistics. It is important to note that, unlike \eqn{cone}, 
this quantity depends on the \emph{strong} convergence properties of the underlying integration scheme \eqn{strong}. In this case, $P_L$ and $P_{L-1}$ must 
be calculated using different timesteps, but the same underlying stochastic path in which the path at the coarser level $L-1$ is suitably compounded from 
those used at the finer level $L$.  Using $V_L$, the number of samples at each level $N_l$ can then be computed according to \eqn{OptNumber} and by noting $V_l = V_L 2^{2\beta(L-l)}$. 

The $l=0$ level is an exception and, analogous to \eqn{VL}, it is given by
\begin{equation}
V_0 \simeq V_0^N \equiv \frac{1}{N} \sum_{j =1}^N \left( P^\rii_0\right)^2 - \frac{1}{N^2} \left(\sum_{\rii=1}^N P^\rii_0  \right)^2,
\label{V0}
\end{equation}
where, again, $N$ should be sufficiently large.

Careful calculation of the constants in this section is essential to obtaining an accurate estimate of $\PLnL$ using the MLMC method\footnote{An 
 alternative approach to bounding the bias error is based on increasing $L$ until the condition $|\dPLN |<\varepsilon/\sqrt{2}$ is met \cite{giles2008multilevel}. In the case of a sign change between successive $\dPLN$, modifications are required. }. (Although it should be noted that even for direct methods, $c_1$ must still be calculated to ensure $\lx{MSE} \leq \varepsilon^2$.) The method can be
implemented, step by step, as follows, where special note should be taken at step 6  where it is essential that each pair of realization be calculated consistently from the
same underlying path. A prescription for doing so is provided in \cite{dimits2013higher}.

The steps are:
\begin{enumerate}
\item Choose a payoff function $P$, end time $T$, and an acceptable error bound $\varepsilon$.
\item  Choose a method of direct integration for the MLMC method.
\item  Use \eqn{cone} to calculate $c_1$, and combine with \eqn{OptLevel} to get $L$.
\item  Use $L$, \eqn{VL} and \eqn{V0} to calculate $V_L$ and $V_0$.
\item Use $V_L$, $V_0$ and \eqn{OptNumber} to calculate $N_l$ and $N_0$.
\item Calculate $N_l$ pairs of  $P_l^\rii, P_{l-1}^\rii$, each with the same underlying stochastic path. 
\item Use $N_l$ pairs of $P_l^\rii, P_{l-1}^\rii$ to calculate $\dPlN$ for each $l=1$ to $L$.
\item Use $N_0$ to calculate $P_0^{N_0}$.
\item Use $P_0^{N_0}$ and $\dPlN$ from $l=1$ to $L$  to calculate $\PLnL$ according to \eqn{PayoffEst}. 
\end{enumerate}

These steps are implemented in Section \ref{Test} for a test case describing the collisional diffusion of a beam of a particles interacting 
with a Maxwellian background.

\section{Beam Diffusion Test Case}\label{Test}

The average pitch-angle evolution of a spatially homogeneous, gyrotropic beam of particles $a$ constitutes a simple and robust test case for the MLMC method. 
The beam is injected into a Maxwellian background distribution of particles $b$ with equal mass, $m_a = m_b$. In the absence of forcing, the action of 
collisions isotropizes $f_a$ in this classical relaxation problem.

Working in spherical velocity-space coordinates $(v, \theta, \phi)$,  $v$ is the particle speed,  $\theta = \cos^{-1} \mu$ is the pitch angle with 
respect to some preferred direction $\mu$, and $\phi$ is the azimuthal angle. Neglecting $\phi$-dependence (i.e. a two-dimensional collision model), the collision operator \eqn{LFP} is \cite{rosenbluth1957fokker}: 
\begin{equation}
\pdif{f_a}{t}\big|_\lx{coll} \!\!= \!\!
-\frac{1}{v^2}\pdif{}{v}\!\!\left[ \left(v^2 \pdif{h}{v} + \pdif{g}{v}\right) \!\! f_a\right] + 
\frac{1}{2 v^2}\frac{\partial^2}{\partial v^2} \left( v^2 \frac{\partial^2 g}{\partial v^2} f_a \right) + 
\frac{1}{2 v^2}\pdif{g}{v}\pdif{}{\mu}\left[ ( 1 - \mu^2) \pdif{f_a}{\mu} \right],
\label{LFP_SPC}
\end{equation}
and the (initial) particle distributions are 
\begin{align}
f_a &= n_a \delta(\vc{v}^* - \vc{v}),\label{initial}\\
f_b &= \frac{n_b}{\left(\pi \vth^2\right)^{3/2}} \exp[-v^2/\vth^2],
\end{align}
where $\vc{v}^*$ is some single valued initial velocity for the test particles and the Maxwellian field particle thermal velocity is $v_{\lx{th},b}^2 = 2 \tau / m_b = (2/3 n_b) \!\int  \! f_b  |\vc{w}|^2 d^3\vc{v}$ where $\tau$ is the temperature of $f_b$ and $\vc{w} = \vc{v} - \vc{u}$ is the random, i.e. particle minus flow, velocity. We set $n_b \gg n_a$ throughout so we can neglect the
back reaction of the beam on the Maxwellian. 

In the case that $f_b$ is Maxwellian,  Trubnikov \cite{trubnikov1965review} gives $g,h$ concisely by:
\begin{align}
g(v) &=   \frac{1}{2}  \Gamma n_b \sqrt{2} \vth \left[ \Phi\left(2x + \frac{1}{x}\right) + \Phi' \right],\label{Max_h}\\
h(v) &= 2 \Gamma  n_b \frac{\Phi}{v}, \label{Max_g}
\end{align}
where $x = v/\vth$, and $\Phi(x)$ is the standard error function. 

\begin{figure}
    \centering
     \includegraphics[width=0.49\textwidth]{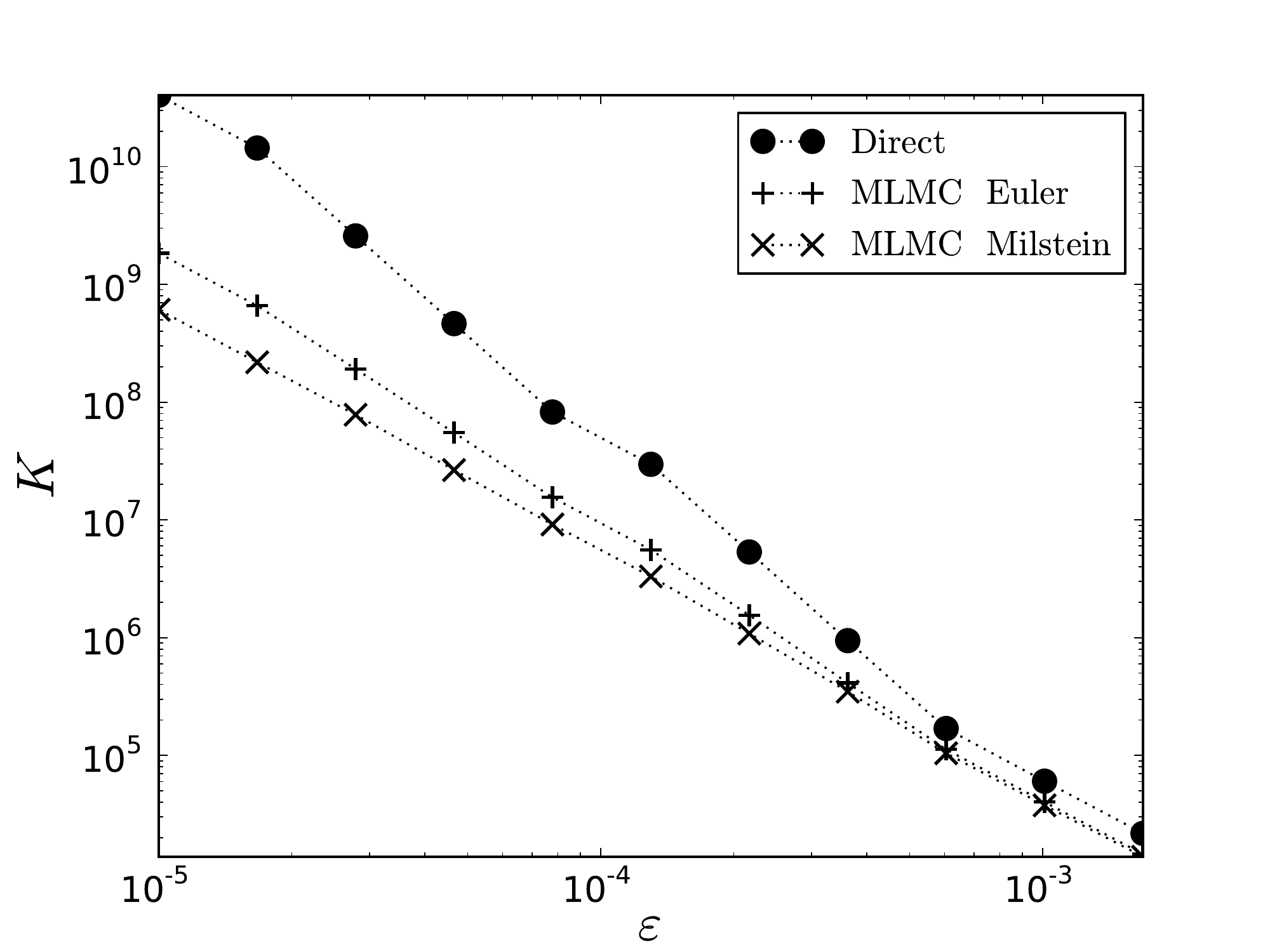}
    \includegraphics[width=0.49\textwidth]{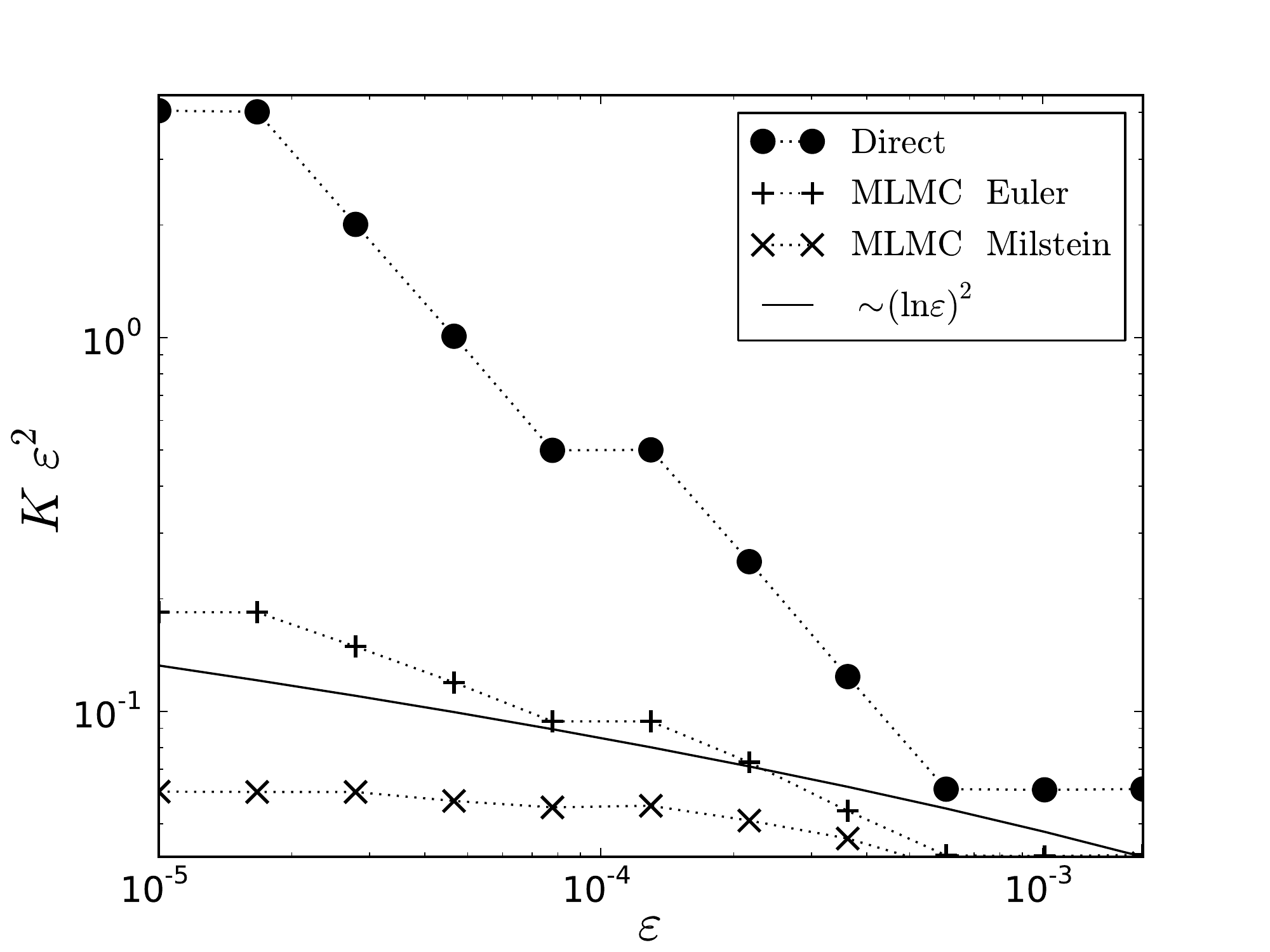}
  \caption{Computational cost $K$ (left) and normalized $K$ (right) versus user-prescribed error bound $\varepsilon$ for the beam diffusion test case in Section \ref{Test}. The parameters are as in Fig. \ref{Fig:Mean}. Both the Euler and Milstein MLMC schemes are more efficient than direct integration, Milstein by a factor of approximately 100 for the case $\varepsilon = 10^{-5}$, and the scaling costs predicted by \eqn{CostSummary} are  recovered. 
  \label{Fig:CostErr2}}
\end{figure}

The set of Langevin equations  \eqn{Langevin} corresponding to  \eqn{LFP_SPC}-\eqn{Max_g} are then given by
\begin{align}
d v(t) &= F(v) dt + \sqrt{D_v (v)} dW_v(t), \label{v_SPC}\\
d \mu(t) &= - 2 D_a(v) \mu dt + \sqrt{2 D_a(v) (1 - \mu^2)} dW_\mu (t)\label{mu_SPC},
\end{align}
where $D_v, D_a$ are the speed and angular diffusion coefficients, and we have normalized $f_a$ by $2 \pi v^2$ as required to bring the derivative to the outside in 
\eqn{LFP_SPC}, so in \eqn{v_SPC} and \eqn{mu_SPC}, $v(t) \to 2\pi v^3(t)$ and $\mu(t) \to 2 \pi \mu(t) v^2(t)$ in un-normalized coordinates. In what follows  we work in normalized coordinates. 

The curvilinear coordinate system requires the coefficients of the deterministic and stochastic terms to be a mixture of $F, D_a, D_v$, and these are given by \cite{dimits2013higher}
\begin{align}
F(v) &= -\frac{A_D}{2 v^2}\left[(4 x + 1) G(x) - \Phi(x)\right],\label{FF}\\
D_v(v) &= \frac{A_D}{2v} G(x),\\
D_a(v) &= \frac{A_D}{4v^3}\left(\Phi(x) - G(x) \right),\label{DADA}
\end{align}
where $A_D= 2 n_b \Gamma = 8 \pi n_b q_a^2 q_b^2 \Lambda/m_a^2$, and $G(x) = (\Phi - x \Phi')/2 x^2$ is the Chandrasekhar function.

The finite-timestep discretized Langevin equations can then be obtained from \eqn{v_SPC} and \eqn{mu_SPC} by an iterative stochastic Taylor expansion in $\sts$. To lowest order, retaining terms up to order $\order(\sts^{1/2})$, we obtain the Euler-Maruyama scheme, and to next order, retaining terms up to $\order(\sts)$, we obtain 
the Milstein scheme. Normalizing time $t$ by the thermal field-particle collision rate $\nu_b = \sqrt{2} A_D/\vth^3$ and velocity $v$ by $\sqrt{2} \vth$, the dimensionless discretized Langevin 
equations are
\begin{align}
\Delta v &= F \sts +\sqrt{2 D_v} \Delta W_v + \kappa_M D'_v \frac{1}{2} \left(\Delta W_v^2 - \sts \right),\label{Deltav_SPC}\\
\Delta \mu &= -2 D_a \mu \sts + \sqrt{2 D_a (1 - \mu^2)} \Delta W_\mu + \nonumber \\ 
&\qquad \qquad \qquad \qquad  \kappa_M \left[ -2 D_a\mu\frac{1}{2} \left(\Delta W_\mu^2 - \sts \right) + \sqrt{\frac{D_v}{D_a}} \sqrt{1- \mu^2} D'_a A_{v\mu}\right],
\label{Deltamu_SPC}\end{align}
where $\kappa_M = 0,1$ for Euler and Milstein respectively and $A_{v\mu}$ is the $v$--$\mu$ correlated random variable \eqn{Levy} whose approximate characteristic function (the Fourier transform of its probability density function) is given in \cite{dimits2013higher}. The coefficient functions are to be evaluated at the start of each timestep, as required by the Ito stochastic calculus, and the normalized drag and diffusion coefficients become
\begin{align}
F(v) &=  2 v (D_a(v) -  D_v(v) ), \label{F_norm}\\
D_v(v) &= \frac{1}{v} G\left(\frac{v}{\sqrt{2}} \right),\\
D_a(v) &= \frac{1}{2v^3}\left[ \Phi\left(\frac{v}{\sqrt{2}}\right) - G\left(\frac{v}{\sqrt{2}}\right)\right].\label{Da_Norm}
\end{align}
In equilibrium, where $f_a$ and $f_b$ have the same flow velocity $\int f_a  \vc{v} \lx{d}^3  \vc{v}/n_a = \int f_b \vc{v} \lx{d}^3\vc{v} /n_b$ and temperature, $(m_a/ 3 n_a)\int  f_a \lx{Trace}(\vc{w w})d^3\vc{v}  = \tau$, the drag and diffusion coefficients are related by the spherical coordinate form of the Einstein relations  \eqn{Einstein}. That is, \eqn{F_norm}-\eqn{Da_Norm} must
obey
\begin{align}
&2 D_a(v) \mu - \frac{\partial}{\partial \mu} \left[D_a(v) (1 - \mu^2) \right] =0,\label{Einstein_1}\\
& F(v) - \frac{1}{v} \frac{\partial}{\partial v} \left( v^2 D_v(v)\right) + v D_v(v) = 0\label{Einstein_2}, 
\end{align}
which can be readily confirmed by direct substitution\footnote{In addition to the Einstein relations, $D_a$ and $D_v$ satisfy $D_v =  d/dv (v^3 D_a)$ which, fundamentally, is a 
consequence of the fact that Coulomb interactions are through a central force and the Landau-Fokker-Planck treatment keeps only small-angle 
scattering interactions.}.

To ensure the coefficients of the discretized Langevin equation are Lipschitz continuous, as required for the MLMC method, \eqn{Deltav_SPC}-\eqn{Da_Norm}
must be numerically regularized upon implementation. The procedure for doing so is described in Appendix \ref{Regularization}.

Equations \eqn{Deltav_SPC}-\eqn{Da_Norm} constitute the building blocks for a MLMC scheme that returns the mean or moment of some payoff $P$ of $\vc{v}$ associated with $f_a$ 
as it interacts with $f_b$. The building blocks are independent of $P$, the time at which its mean is evaluated $T$, and the acceptable error bound $\varepsilon$. 
These quantities are parameters of the simulation. Collectively they determine the preconditioning parameters of the method, $c_1, V_L$ and $V_0$.  

We choose our payoff function  
\begin{equation}
P  = \mu, \nonumber 
\end{equation}
so that the MLMC scheme approximates its mean value over all particles 
\begin{equation}
\hat{P}(T) \simeq \PLnL(T)  \simeq \frac{1}{n_a} \int  \mu f_a(T) d^3 \vc{v}.
\label{mu_payoff}
\end{equation}

\begin{figure}
    \centering
    \includegraphics[width=0.49\textwidth]{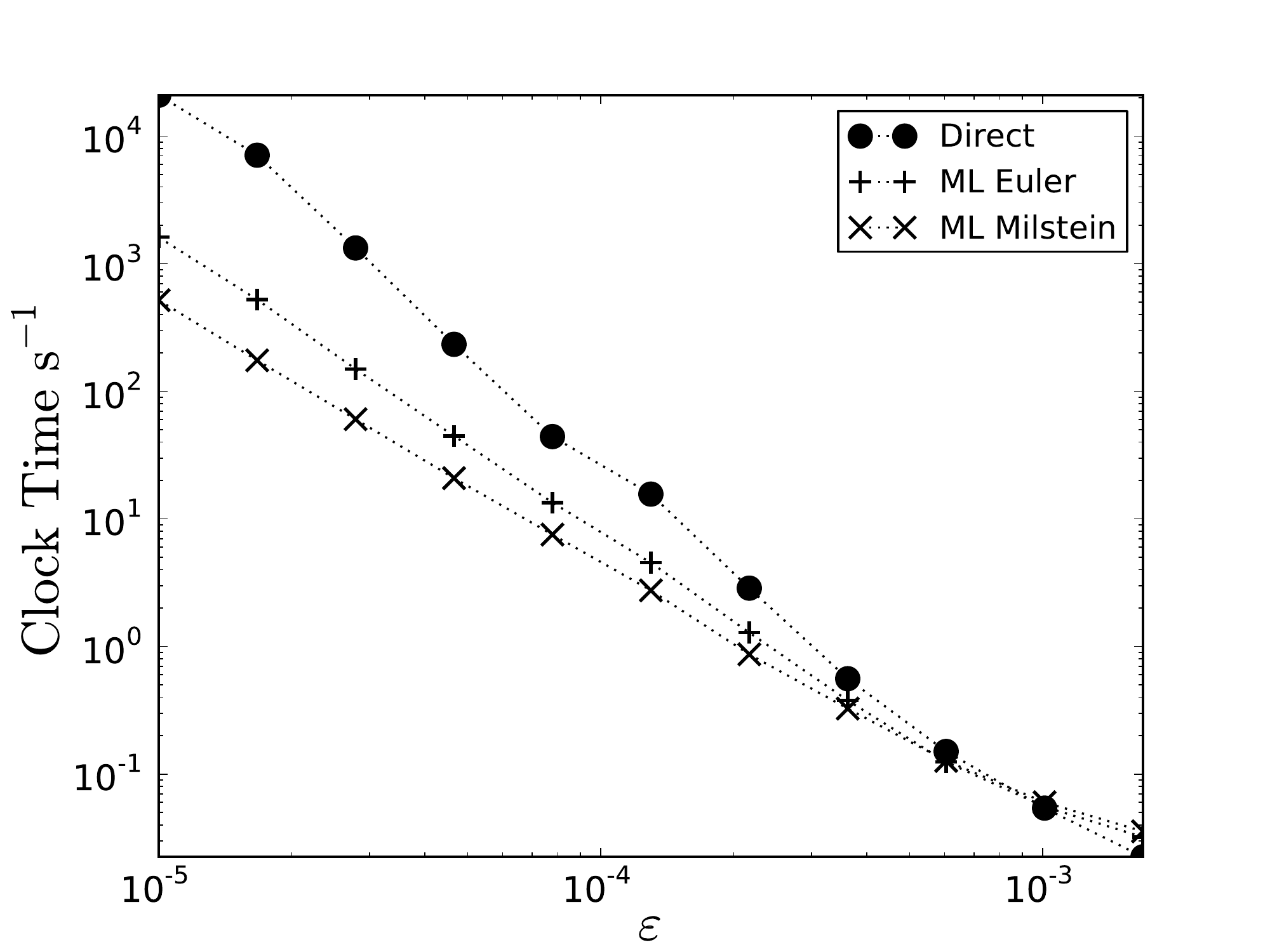}
  \caption{Wall clock time to execute steps 1-9 of Section \ref{Implementation} versus user prescribed error bound $\varepsilon$ for the same parameters as in Fig. \ref{Fig:Mean}.   The numerical code is written in Python and Fortran 90, 
  and executed on a 2.4 GHz Intel Core i5 MacBook.  The MLMC methods are significantly faster than the direct methods for high accuracy simulations. For small values of $\varepsilon = 10^{-5}$, the Milstein method is approximately $40$ times faster than the direct alternative. }
\label{Fig:ClockTime}\end{figure}

\begin{figure}
    \centering
    \includegraphics[width=0.49\textwidth]{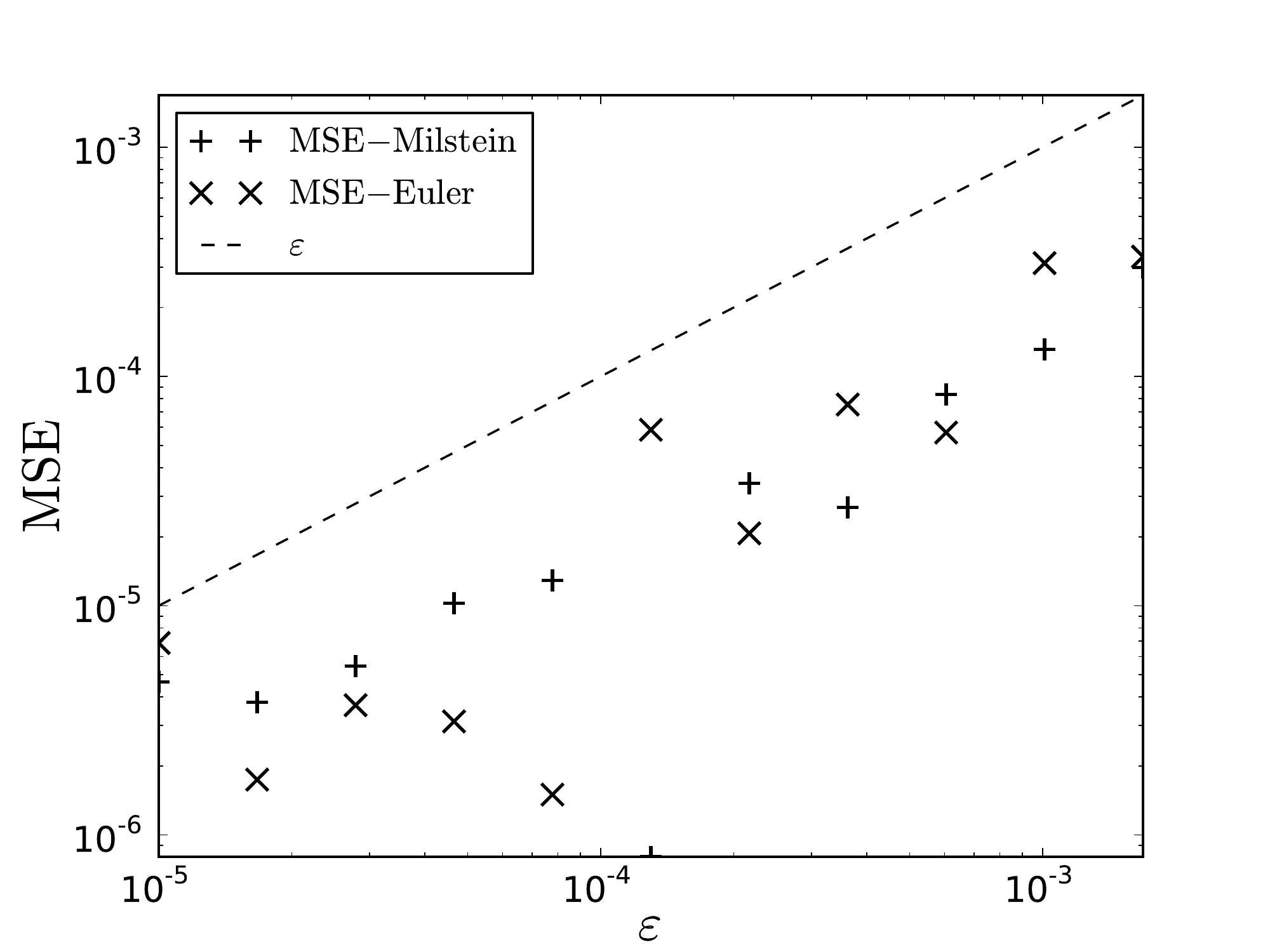}
  \caption{Mean Squared Error  ($\lx{MSE}$) vs user prescribed error bound $\varepsilon$ for a range of MLMC runs with both Euler and Milstein \eqn{MSEDfn2}. The MLMC method accurately approximates the true payoff to within $\varepsilon$, as all points fall below the delimiting dashed line. Parameters are as in
  Fig \ref{Fig:Mean} and the mean is taken over 10 independent MLMC runs. 
  \label{Fig:CostErr}}
\end{figure}

The results of numerical implementation are shown in Figs. \ref{Fig:CostErr2}, \ref{Fig:ClockTime} and \ref{Fig:CostErr}  where the method successfully approximates the right hand side of \eqn{mu_payoff}, the `true' value of which is itself approximated  using a high-resolution many particle direct simulation Monte-Carlo scheme (the direct Euler scheme with $2 \cdot 10^9$ particles and $2^8$ timesteps.). For the 
most accurate case $\varepsilon = 10^{-5}$, both MLMC methods are considerably faster than the direct method for which the parameters (timestep, sample size) are chosen such that the $\lx{MSE}\leq\varepsilon^2$. The Milstein MLMC method is approximately 100 times faster than the direct method  in terms of its computational cost, Fig. \ref{Fig:CostErr2}, and 40 times faster in terms of its wall clock timing, i.e. the number of seconds required to complete the computation on a computer, Fig. \ref{Fig:ClockTime}. 

The difference between the computational cost and wall clock timing arises, in part, because the MLMC method actually performs two integrations at each level $l$, a coarse and a fine integration. This leads to an additional cost of $3/2$ not captured by \eqn{CompCost2}. Furthermore, the MLMC Milstein method contains additional terms that include the L\'evy areas. These must be calculated at an additional cost relative to the direct and Euler MLMC methods.

The MLMC scheme accurately (to within  $\varepsilon$) describes the average pitch angle relaxation of a beam
of particles interacting with an isotropic Maxwellian background. The scaling predictions given by \eqn{CostSummary} are reproduced, so that the computational cost (total number of operations) of the Milstein, Euler and direct methods scale as $\varepsilon^{-2}, (\ln \varepsilon)^2\varepsilon^{-2}$, and $\varepsilon^{-3}$ respectively. Integral to the Milstein MLMC method for $P = \mu$ is an accurate description of $A_{v\mu}$, \eqn{Deltamu_SPC}.  For other payoffs in two dimensions that are independent of $\mu$, cross-terms like $A_{v\mu}$ are not required. 
For example, 
\begin{equation}
P = \frac{1}{2} m_a v^2 \nonumber
\end{equation}
approximates the mean kinetic energy per particle $\mathcal{K}$
\begin{equation}
\hat{P}(T) \simeq \PLnL(T)  \simeq \frac{1}{2} \frac{m_a}{n_a} \int f_a(T, \vc{v} ) |\vc{v}|^2 d^3\vc{v}, \nonumber
\end{equation}
where the underlying Milstein scheme includes additional terms proportional to $\Delta W_v^2$ and $\sts$ only.

\section{Discussion}\label{Discussion}

The MLMC method constitutes a powerful new technique for solving kinetic expectation problems, the solutions to which can be used to reconstruct the 
underlying distribution function. Asymptotically, it has significantly improved scaling properties compared to both direct SDE and binary collision methods. However, the method is  limited in several respects. 

Firstly, for any given problem, the multiplicative constant associated with the scaling of the computational cost may make the method prohibitively 
expensive. Secondly, it is unclear how, for $t<T$, mean field quantities 
like electromagnetic fields and evolving back-reacted drag and diffusion coefficients that dictate particle trajectories, are to be computed. Thirdly, for strongly non-equilibrium 
 problems, a large number of moments or binning operations
 may be needed to accurately reconstruct $f$. This could be expensive. 
 
 In this section we discuss techniques and extensions to the MLMC method that can be used to address these problems. 

\subsection{Improved efficiency}

For expediency, our description of the MLMC method presented in Section \ref{sec:MLMC} was basic and concise. However, several 
improvements, not described earlier, exist.

  Throughout, we have set the refinement factor between levels $M = \sts/\Delta t_{l+1} = 2$. 
Giles \cite{giles2008multilevel} has argued that  while this choice is optimal for multilevel elliptic PDE solvers, for the MLMC method, other values may 
improve efficiency -  specifically, a factor two saving may be obtained by setting $M = 7$ in the MLMC Euler scheme. Less extreme values, $M = 3,4$ etc, also 
lead to improvements. Similarly, multiplicative constants other than a half for the finite timestep and finite sampling bounds, $\varepsilon^2/2$,
in \eqn{condition} may also lead to improvements. 

Adaptive algorithms, and quasi-Monte Carlo sampling can also increase the efficiency of the method \cite{hoel2012adaptive, giles2009multilevel, 
gerstner2013randomized}. In particular, for the Milstein method, the rapid diminishing of the finite-sampling error with decreasing $\sts$ means that the 
vast majority of the computational effort is expended at the coarsest $l=0$ level - Figure \ref{Fig:Nl}. Quasi-Monte Carlo methods are well suited to reducing 
this cost.  Moreover, when the function $P$ is sufficiently smooth (twice differentiable) - as would be expected in many plasma physics applications - the cost of simulating the coarsest level may be completely eliminated using the recently developed Ito linearization technique \cite{ricketson2013MLMC}.

For existing code bases that implement the direct Euler scheme, it is possible to obtain the optimal $\varepsilon^{-2}$ scaling of the MLMC scheme without 
simulating the L\'evy areas. Using antithetic techniques, the sum in \eqn{MLMCOpt} can be bound, even though the underlying integrator has $\beta = 1/2$ \cite{giles2012antithetic}. Generalizations of the antithetic method exist \cite{ricketson2013MLMC, Giles_ext_antithetic2013}.

Finally, because sampled paths are independent, like direct simulation Monte Carlo, the method can be readily parallelized. Indeed, the timescale over
which the sampled paths of \eqn{Langevin} can be integrated independently is only limited by the requirement that the mean fields are updated. 

\subsection{Mean fields}

When the macroscopic dynamics of a system and its collisions occur on the same timescale,  direct simulation Monte Carlo and particle-in-cell codes must perform two operations at each numerical timestep \cite{qiang2000self}. The first, is to advance 
the particles' positions $\vc{x}$ and velocities $\vc{v}$ according to the discretized equations of motion, including the mean force fields and collisions. In Langevin form,
including spatial dependence and electromagnetic forces, the equation of motion \eqn{Langevin} generalizes to \cite{hirvijoki2013monte}:
\begin{align}
\lx{d}\vsb{x}{\cii} &= v_\cii dt \label{dx}\\
\lx{d}\vsb{v}{\cii} &= \left\{\frac{e}{m} \left[ \vsb{E}{\cii} +  c^{-1}  \epsilon_{\cii \cjj \ckk} \vsb{v}{\cjj} \vsb{B}{\ckk} \right]  + \vsb{F}{\cii}\right\} dt + \vsb{D}{\cii \cjj} d\vsb{W}{\cjj},\label{dv_general}
\end{align}
where $c$ is the speed of light, $\epsilon_{\cii\cjj\ckk}$ is the Levi-Civita symbol, and $E_{\cii}, B_{\cii}$ are the mean electric and magnetic fields.
To ensure particle trajectories are calculated accurately over an extended period, a second operation must then be performed - the mean fields must be updated based on the new particle
positions and velocities. 

The mean electromagnetic fields, $E_\cii, B_\cii$ are functions of $t,\vc{x}$ only. They can be calculated from Maxwell's equations, a set 
of coupled first order linear PDEs, in terms of the sum over species of the macroscopic charge density $\rho = e n$ and current $J_{\cii} = e n u_{\cii}$ at each point in space. 

In the Langevin framework, the macroscopic quantities can be accurately calculated on a timescale $T$ using the MLMC scheme and an appropriate choice of payoff. For 
$\rho$, the payoff is 
\begin{align}
P &= e n \Theta(\vc{x}'),
\label{DensityPayoff}
\end{align}
where 
\begin{equation} 
\Theta = \left\{ \begin{aligned}
&1, & \quad \vc{x}' < \vc{x} < \vc{x}'+ \delta \vc{x}, \\
&0, & \quad \vc{x}'  > \vc{x} > \vc{x}' + \delta \vc{x}, 
\label{binning}
\end{aligned} 
\right. 
\end{equation}
is a binning function\footnote{ Formally, the MLMC method requires $P$ to be Lipschitz, and so simple step functions are inappropriate.  
Modifications to $\Theta$ to ensure Lipschitz continuity may be necessary, but, nevertheless, improved efficiency relative to direct methods has been shown for a variety of non-Lipschitz payoffs \cite{giles2009analysing}.} for the real-space grid cell at position $\vc{x}'$ and of size $\delta \vc{x}$, and $n$ is the initial macroscopic density, as in \eqn{initial}. It follows that $\rho(T, \vc{x}')$ is  given by 
\begin{align}
\hat{P}(T) \simeq \PLnL(T)  \simeq  e \int f(T, \vc{x}', \vc{v}' ) d^3\vc{v}'
\label{density_payoff}
\end{align} 
and similarly for $J_\cii$ with the payoff $P = e n \Theta(\vc{x}') v_\cii$.  

Along with Maxwell's equations and the equations for $F, D$ in section \ref{Formulation},  \eqn{dx}-\eqn{density_payoff} provides a complete, efficient 
plasma description using MLMC methods. The description is, however, only efficient \emph{and} accurate when it is acceptable to resolve the coefficients of 
\eqn{dv_general}, including $F$ and $D$, self-consistently, on a slow timescale $T$. That is, if the inherent timescale on which the macroscopic mean fields evolve is $\ll T$, the MLMC method
will fail to capture important dynamics that take place on this faster timescale. In its present form, the method is therefore restricted to the small Knudsen number regime where 
the collisional dynamics occur on a faster timescale than the macroscopic dynamics\footnote{The method can also be applied when the Lorentz force is absent or externally imposed, and $n_t \ll n_f$ or the collisions are between electrons and ion. In this case, the back reaction of the test particles can also be neglected.}.  Within this
framework,  the MLMC method could itself constitute a building block for a multiscale simulation in which the collisions and macroscopic 
dynamics are resolved on timescales of $\order(\sts)$ and $\order(T)$ respectively.   

It remains an open challenge to extend the MLMC method to kinetic problems in which there is no clear scale separation between the collisional and 
macroscopic dynamics i.e. Knudsen numbers of order one and greater. However, if this challenge could be met, it would constitute a potential game changer for  kinetic plasma simulations in general.

\subsection{Distribution functions}\label{Sec:Distribution}

Thermalized distributions vary smoothly on a velocity-space scale $v_{\lx{th}}$ and can be uniquely determined from their first three moments. For non-equilibrium distributions, this is not the case. Two simple methods for reconstructing non-thermal $f(T)$ using the MLMC method exist. 

The first is by summing the moment hierarchy, where each moment is calculated using the MLMC method with an appropriate choice of payoff $P$. For a complete set of moments,  
the structure of $f$ can be captured identically (Chapter 7, \cite{cercignani1988boltzmann}). For a finite-subset, as is practically achievable,  an accurate  approximation can still be obtained \cite{hammett1990fluid}. 

The second method for determining $f$ is through a generalized version of \eqn{DensityPayoff} that bins particles in both real and velocity space\footnote{This method is qualitatively similar to an inverse semi-Lagrangian process \cite{sonnendrucker1999semi}, and comes at an informational cost equivalent 
to that incurred in a resampling procedure. }. In this  case $P = n \Theta(\vc{x}', \vc{v}')$ and  $\Theta$ is a simple generalization of \eqn{binning} to include velocity space cells $\vc{v}'$ of size $\delta \vc{v}$.
It follows that  $\hat{P}(T) \simeq \PLnL(T)  \simeq   f(T, \vc{x}', \vc{v}' )$ returns the particle density in a phase space cell $\vc{x}', \vc{v}'$ at time $T$.

Returning multiple outputs from a single
 run is useful for both the methods above \cite{giles2013review}. Statistical errors withstanding, the same set of paths is needed to compute both successive moments and the binned phase-space distribution. So, by storing and re-using paths, the computational cost of calculating multiple moments is only approximately as much as the most expensive moment. The same is true for phase-space binning. 
 
So far our discussion has focused on calculating distributions at time $T$. While extending the method to multi-valued initial conditions, 
unlike those in Section \ref{Test}, is not technically difficult, a number a comments are in order. 

First, chaotic particle trajectories, real (e.g. tokamak wall) and velocity space (e.g. magnetic mirror) boundaries are ubiquitous in plasma physics. Particles with nearby initial conditions, sampled from the 
same spatial cell, may drastically diverge in phase space. The consequences of this for the MLMC method are unknown.

Second, multi-valued initial conditions introduce a second source of statistical error, beyond that attributable to Brownian motion. When the variance
in the initial data is much less than that associated with the random walk $\Var[v(0)] \ll \Var[v(T)]$, the computational cost of the simulation is unchanged. However, when 
the converse is true, the cost may increase dramatically.  The ratio of the two terms is approximately
\begin{equation}
\frac{\Var[v(T)}{\Var[v(0)]} \sim \frac{T e^4}{m_a^2} \frac{n_b v_{\lx{th},b}^2}{n_a v_{\lx{th},a}^2},\nonumber
\end{equation}
where we have assumed that the initial conditions for $v(T)$ in the numerator are single-valued, i.e. given by \eqn{initial}, and that $T \ll$ the macroscopic timescale. 

\section{Conclusion}\label{Conclusion}

For the first time, we have shown how the multilevel Monte Carlo integration scheme can be used to simulate Coulomb collisions in a plasma. Asymptotically, the method is up to $\varepsilon^{-1}$ times faster than standard direct simulation Monte Carlo or binary collision methods, when used with
an underlying Milstein discretization. This is illustrated in Fig. \ref{Fig:CostErr2} where the total computational cost (operations count) for the direct SDE and the Euler and Milstein multilevel schemes are shown to scale as predicted. We have also demonstrated that the multilevel schemes are significantly faster than direct SDE methods  in terms of both computational cost, Fig. \ref{Fig:CostErr2}, and wall clock time, Fig. \ref{Fig:ClockTime}.  Our numerical results are for a classic beam diffusion test case in 2$D$ and over a given range of prescribed errors

The most important extension to this work would be an expansion of the method to arbitrary Knudsen number problems, where a separation 
of collisional and macroscopic timescales does not exist. Other valuable studies would also include a demonstration of the method in forced, spatially inhomogenous and multi-physics problems, and an extension to kinetic collisions models other than the Coulomb case, for example neutral particle collisions.  

\subsubsection*{Acknowledgements}

Special thanks to the LLNL Visiting Scientist Program for hosting M.S.R. and L.F.R., and to B. Albright, S. Brunner, A. Cerfon, L. Chac\'on, F. Fi\'uza, M. Giles, M. Landreman,  T. Wood, and B.Yan for helpful discussion throughout. This work was performed by UCLA under Grant DE-FG02-05ER25710 and by LLNL under Contract DE-AC52Ð07NA27344 under the auspices of the US Department of Energy Advanced Scientific Computing Research's Multiscale Mathematics Initiative.

\newpage
\appendix

\section{Numerical Regularization}\label{Regularization}

A number of numerical and modeling poles must be circumnavigated to implement the MLMC method successfully.

\subsubsection*{Diffusion to negative speeds}

The Langevin equation governing the evolution of $v$ is \eqn{Deltav_SPC}. Finite changes in $\Delta v$ arising from terms containing  $\Delta W_v$ can be of any size, although large values are (exponentially) unlikely. It follows that  $v(t + \sts) = v(t) + \Delta v$ can be such that $ v(t + \sts) \leq 0$. This is not only unphysical, but also numerically problematic as $F, D_a$ become singular at a rate $v^{-1}, v^{-2}$ respectively as $v \to 0^+$. Furthermore, the deterministic drag coefficient $F$ is anti-symmetric in $v$, so if $v<0$ the first term in \eqn{Deltav_SPC} drives $v$ to yet more negative values. 

Our approach is to regularize the coefficient of the  deterministic drag term $F$ in \eqn{Deltav_SPC}, and the stochastic diffusion term $D_a$ in \eqn{Deltamu_SPC}. Our method  differs from that of Lemons et al. \cite{lemons2009small} who did not  account for the small, but finite, probability case that particles diffuse to $v<0$, even when the coefficients of the diffusion terms are set to zero for small $v$. We define the piecewise Lipschitz continuous functions
 \begin{equation} 
\mathcal{F}(v) = \left\{ \begin{aligned}
& F(v) , & \quad v > v_c \\
& \frac{F'(v_c)}{2 v_c} (v^2- v_c^2) + F(v_c), & v \leq v_c\end{aligned} 
\right.,\label{Reg_F}
\end{equation}
and
\begin{equation} 
\mathcal{D}_a(v) = \left\{ \begin{aligned}
& D_a(v) , & \quad v > v_c \\
& \frac{D'_a(v_c)}{2 v_c} (v^2- v_c^2) + D_a(v_c), & v \leq v_c\end{aligned} 
\right.,\label{Reg_Da}
\end{equation}
where $v_c$ is the critical value of $v$ at which regularization occurs, and $F', D_a' = \lx{d} D_a/\lx{d} v, \lx{d} F/\lx{d} v$. 

Direct substitution of \eqn{Reg_F} into \eqn{Deltav_SPC} yields a regularized equation for $\Delta v$:
\begin{equation}
\Delta v = \mathcal{F} \sts +\sqrt{2 D_v} \Delta W_v + \kappa_M D'_v \frac{1}{2} \left(\Delta W_v^2 - \sts \right).
\end{equation}
An analogous modification of \eqn{Deltamu_SPC} also follows, but further regularization of the imaginary diffusion coefficients is first required. We note that in the small region $v<v_c$, the Einstein relation \eqn{Einstein_2} is not obeyed. In the simulations conducted here we set $v_c = 0.05$, which we find, empirically, to work.

\subsubsection*{Imaginary coefficients}

The Langevin equation governing the evolution of  $\mu$ is  \eqn{Deltamu_SPC}. Analogous to the previous section, finite changes in $\Delta \mu$ driven by large values of $\Delta W_\mu, A_{v\mu}$ can results in $\mu(t+\sts) = \mu(t) + \Delta \mu$ being  such that  $|\mu| >1$. It follows that $\sqrt{1-\mu^2}$ can become imaginary, which is unphysical. 

 To constrain the discretized equations to be physical, we define the modified coefficient  $\mm$ to be:
 \begin{equation}
\mm(\mu) = \left\{ \begin{aligned}
&\sqrt{1-\mu^2} & \quad |\mu| < \mu_c\\
&\sqrt{1-\mu_c^2}\exp [ (\mu- \mu_c) S(\mu_c) ] & \quad |\mu| \geq \mu_c\\
\end{aligned} 
\right. ,\label{Reg_mu}
\end{equation}
where $S(\mu_c) = \mu_c/(1-\mu_c^2)$, and $\mu_c$ is the critical value at which regularization occurs. 

The coefficient is  unaltered away from the critical poles, and regularized near them when $|\mu| > \mu_c$ in a manner consistent with the
Einstein relation \eqn{Einstein_1} and the condition that the coefficients are Lipschitz continuity. 

Substituting \eqn{Reg_Da} and \eqn{Reg_mu}   into \eqn{Deltamu_SPC}, the regularized evolution equation for $\Delta \mu$ is 
\begin{equation}
\Delta \mu = 2 \mathcal{D}_a \mm \mm'  \sts+ \sqrt{2 \mathcal{D}_a \mm } \Delta W_\mu + \kappa_M \left[  \mathcal{D}_a\mm \mm' \left(\Delta W_\mu^2 - \sts \right) + \sqrt{\frac{D_v}{\mathcal{D}_a}} \mm \mathcal{D}'_a A_{v\mu}\right],
 \end{equation}
where $\mm' = \lx{d} \mm/\lx{d} \mu$. In the simulations conducted here we set $\mu_c = 0.95$ which, again, we find to work empirically.

\bibliographystyle{unsrt}	%
\newcommand{\mnras}[0]{MNRAS}\newcommand{\apj}[0]{ApJ}\newcommand{\apjs}[0]{ApJ}\newcommand{\apjl}[0]{ApJ
  Letters}\newcommand{\araa}[0]{ARA\&A}\newcommand{\aap}[0]{A\&A}

\end{document}